\documentclass[aps,preprint,pre,superscriptaddress]{revtex4-1}

\usepackage[utf8]{inputenc}
\usepackage{float}
\usepackage[fleqn]{amsmath}
\usepackage[title]{appendix}
\usepackage{amsmath,geometry,amssymb,verbatim,graphicx,caption,makecell,subcaption,tikz,color,hyperref,float}
\usepackage{geometry}
\usepackage{enumitem}
\usepackage{siunitx}
\def\ba{\begin{eqnarray}}
\def\ea{\end{eqnarray}}
\def\be{\begin{equation}}
\def\ee{\end{equation}}
\def\bm{\begin{math}}
\def\me{\end{math}}

\begin{document}
	
\title{Domain Growth and Aging in the Random Field XY Model: A Monte Carlo Study}
	
\author{Ramgopal Agrawal}
\email[]{ramgopal.sps@gmail.com}
\affiliation{School of Physical Sciences, Jawaharlal Nehru University, New Delhi -- 110067, India.}
\author{Manoj Kumar}
\email[]{manojkmr8788@gmail.com}
\affiliation{Centre for Fluid and Complex Systems, Coventry University, Coventry CV1 5FB, UK.}
\author{Sanjay Puri}
\email[]{purijnu@gmail.com}
\affiliation{School of Physical Sciences, Jawaharlal Nehru University, New Delhi -- 110067, India.}
	
\begin{abstract}
We use large-scale Monte Carlo simulations to obtain comprehensive results for domain growth and aging in the random field XY model in dimensions $d=2,3$. After a deep quench from the paramagnetic phase, the system orders locally via annihilation of topological defects, i.e., vortices and anti-vortices. The evolution morphology of the system is characterized by the correlation function and the structure factor of the magnetization field. We find that these quantities obey dynamical scaling, and their scaling function is independent of the disorder strength $\Delta$.  However, the scaling form of the autocorrelation function is found to be dependent on $\Delta$, i.e., {\it superuniversality} is violated. The large-$t$ behavior of the autocorrelation function is explored by studying aging and autocorrelation exponents. We also investigate the characteristic growth law $L(t,\Delta)$ in $d=2,3$, which shows an asymptotic logarithmic behavior: $L(t,\Delta) \sim \Delta^{-\varphi} (\ln t)^{1/\psi}$, with exponents $\varphi, \psi > 0$.
\end{abstract}

\maketitle

\newpage
\section{Introduction}
\label{intro}

The equilibrium properties of spin systems with continuous symmetry (comprised of $n$-component spins) and quenched disorder have received some attention over the years. They have been proposed and studied as models for various experimental systems, e.g., vortex lattices in type-II superconductors \cite{larkin1979pinning,larkin1970effect,nattermann2000vortex}, charge-density waves in the presence of random pinning \cite{fukuyama1978dynamics,lee1979electric,sham1976effect,efetov1977charge}, liquid crystals in porous media \cite{iannacchione1993randomly}, amorphous ferromagnets with random anisotropy \cite{sellmyer1992random}, etc. These systems are highly sensitive to quenched disorder in low dimensions. Thus, interest has focused on how disorder affects the phase transitions in the corresponding disorder-free systems.

A text-book example of spin systems is the Ising model, where the spins sitting at lattice sites can take discrete values $+1$ or $-1$, and the Hamiltonian possesses a $\mathbb{Z}_2$ symmetry. In the absence of disorder, the Ising model in dimension $d \ge 2$ shows a second-order phase transition at non-zero temperature, i.e., ferromagnetic {\it long-range order} (LRO) exists below a critical temperature. The generalization of the Ising model in terms of vector spins ($n > 1$) is invariant under a $\mathcal{O}(n)$ transformation, e.g., XY ($n=2$) model and Heisenberg ($n=3$) model. In the XY model, the spins are free to take any value on a circle. In the Heisenberg model, the spins can take values on the surface of a sphere. In both systems, a low-temperature LRO phase arises only for $d \ge 3$. However, the XY model in $d=2$ exhibits a special phase transition known as the {\it Berezinskii-Kosterlitz-Thouless} (BKT) transition at non-zero temperature $T_{\rm BKT}$ \cite{berezinskii1971destruction,kosterlitz1973ordering,kosterlitz1974critical}. For $T < T_{\rm BKT}$, the system exhibits a phase with so-called {\it quasi-long-range order} (QLRO).

The presence of quenched randomness in terms of an on-site random pinning field has strong effects on the low-temperature properties of the above systems. In the Ising model with random field disorder, no LRO exists at a non-zero temperature in dimension $d \leq 2$. On the other hand, for random-field problems with $n>1$, the heuristic arguments of Imry and Ma \cite{imry1975random}, and rigorous analysis \cite{aizenman1989rounding,*aizenman1990rounding} have shown that there is no LRO in dimension $d<4$. In disordered XY models, it is natural to ask whether there is any low-temperature QLRO phase analogous to the $d=2$ pure XY system.

In particular, for XY systems with a quenched random field, i.e., the random field XY model (RFXYM) as defined below, this question has been the subject of considerable discussion. Studies based on replica-symmetry breaking arguments \cite{nattermann2000vortex,korshunov1993replica} and other theoretical approaches \cite{garel1996variational,feldman2001quasi,giamarchi1995elastic,fisher1997stability} have argued for the existence of QLRO at low disorder in the form of a {\it Bragg glass} phase. Numerical studies \cite{gingras1996topological,fisch1997power,fisch2000random,fisch2007structure} predicted a similar topological phase transition to a {\it pinned vortex free} phase at a non-zero critical field strength $\Delta_c(T)$ in $d=3$ RFXY systems. In $d=2$, this phase was shown to occur only at $\Delta_c = 0$. The work of Tissier and Tarjus \cite{tissier2006unified,tissier2006two,tarjus2020random} contradicts the above predictions. Using the functional renormalization group approach, they claim that the lower critical dimension $d_c^l $ for the QLRO phase is 3.9, i.e., there is no QLRO in $d=2,3$ RFXY systems.

The RFXYM \cite{gingras1996topological,fisch1997power,fisch2000random,fisch2007structure,garanin2013random} is a disordered spin model, where a two-component vector spin $\mathbf{S}_i$ and quenched (static) random field $\boldsymbol {\epsilon}_i$ are placed at each site $i$ of the lattice. The Hamiltonian for the RFXYM with $N$ spins is given as
\ba
\label{eq1}
\mathcal H &=& -J \sum_{\langle ij \rangle} \mathbf{S}_i \cdot \mathbf{S}_j - \Delta \sum_{i=1}^N \boldsymbol {\epsilon}_i \cdot \mathbf{S}_i \nonumber \\
&=& -J \sum_{\langle ij \rangle} \cos(\theta_i-\theta_j) - \Delta \sum_{i=1}^N \cos(\phi_i-\theta_i) .
\ea
Here, each spin $\mathbf{S}_i = (\cos \theta_i,\sin \theta_i)$ and random field $\boldsymbol {\epsilon}_i = (\cos \phi_i,\sin \phi_i)$ are unit vectors, described by the angle variables $\theta_i$ and $\phi_i$, respectively. The strength of the ferromagnetic exchange coupling is $J>0$. The subscript $\langle ij \rangle$ denotes a sum over all nearest-neighbor pairs. The quenched random field variables $\{\phi_i\}$ are drawn uniformly in the interval (0, $2\pi$), and the parameter $\Delta$ defines the disorder strength.

Depending upon how quenched disorder couples to an XY system, there are several versions of the disordered XY model, such as the random-bond XY model (RBXYM)
\cite{ray1992chirality,tang2015thermodynamics,kumar2017ordering}, random-phase XY model \cite{li1996vortex}, site-diluted XY model \cite{leonel2003monte,wysin2005extinction}, or bond-diluted XY model \cite{surungan2005kosterlitz}. The phase transitions in these variants are reasonably well-understood in lower dimensions. The case of the RFXYM is quite different. After several decades of intense investigation, the low-temperature ordered phase in the RFXYM still remains a mystery.

After a deep quench from the high-temperature paramagnetic phase, spin systems locally order via annihilation of topological defects in order to minimize their free energy. The coarsening process is characterized by a time-dependent length scale $L(t)$, which is the {\it domain size} or the {\it inverse defect density}. In scattering experiments, this can be obtained as the inverse peak of the time-dependent structure factor. In simulations, it is usually calculated as the typical decay scale of the order-parameter correlation function. If the system exhibits LRO in its low-temperature equilibrium state, the ordering length scale $L(t) \rightarrow \infty$ for infinite system size. On the other hand, in the absence of LRO, the equilibration proceeds till $L(t)$ becomes comparable to $\xi (T)$, where $\xi (T)$ is the equilibrium correlation length at quench temperature $T$. Recall that, in the Ising system, $\xi (T)$ measures the thickness of domain walls between bulk domains and diverges at the critical point \cite{tcf10}.

In the presence of random external fields, defects are locally pinned, resulting in slow dynamics of the system \cite{sp04}. In this paper, we study the ordering dynamics of the RFXYM for $d<4$ after an instantaneous quench. In the pure $d=2$ XY model, $L(t)\sim (t/\ln t)^{1/2}$  \cite{yurke1993coarsening}. In $d=3$, the XY growth law is the conventional Lifshitz-Allen-Cahn (LAC) law, $L(t)\sim t^{1/2}$ \cite{kohring1986role,gottlob1993critical}. In this paper, we investigate the nature of the asymptotic growth law in the presence of quenched disorder. Apart from the growth laws, the ordering kinetics of spin systems has many other interesting characteristics, e.g., dynamical scaling of the correlation function and structure factor, aging phenomena in two-time quantities, etc. The inclusion of quenched disorder in the system raises several important questions, e.g., robustness of the scaling functions, effects on aging behavior, etc.

The major observations of our present study are as follows. \\
(1) The correlation function and structure factor of the magnetization field exhibit dynamical scaling at all disorder values in $d=2,3$. Further, the scaling functions are found to be independent of disorder. \\
(2) The autocorrelation functions scale for all values of disorder. However, the scaling functions depend on the disorder amplitude, showing the violation of superuniversality. \\
(3) The asymptotic growth law $L(t)$ is logarithmic in $d = 2,3$.

This paper is organized as follows. In Sec.~\ref{method}, we present simulation details. We also discuss the framework for studying domain growth in disordered systems. In Sec.~\ref{simulations}, we present detailed numerical results for $d=2,3$. Finally, in Sec.~\ref{summary}, we conclude this paper  with a summary and discussion of our findings.

\section{Simulation details and methodology}
\label{method}

We study ordering and aging in the RFXYM on cubic lattices in $d=2, 3$ with periodic boundary conditions in all directions. Initially, the system is  prepared in a high-temperature disordered phase by assigning random initial orientations to each spin $\theta_i \in (0, 2\pi)$. The critical temperatures (in units of $J/k_B$) for a pure XY system are $T_{BKT}\simeq 0.89$ in $d=2$ \cite{tobochnik1979monte,fernandez1986critical}; and $T_c \simeq 2.202$ in $d=3$ \cite{hasenbusch1990critical}. At time $t=0$, we rapidly quench the RFXYM to $T=0.2$ in $d=2$, and $T=0.5$ in $d=3$. We evolve the system up to $t=10^6$ Monte-Carlo steps (MCS) via the conventional Metropolis algorithm \cite{newman1999monte} with non-conserved kinetics. A randomly chosen spin $\mathbf S_i$ is given a small random rotation $\delta \in (-0.1, 0.1)$ to a state $\mathbf S_i'$. The new state with angle $\theta_i' = \theta_i + \delta$ is accepted with the Metropolis transition probability $W=\min[1,\exp(-\delta \mathcal{H}/k_B T)]$. Here, $\delta \mathcal{H}$ refers to the change in energy resulting from the change $\mathbf S_i \to \mathbf S_i'$, i.e.,
\begin{equation}
\delta \mathcal H = (\mathbf S_i-\mathbf S_i')\cdot \left(J \sum_{L_i} \mathbf S_{L_i} +\Delta \boldsymbol \epsilon_i \right) ,
\end{equation}
where $L_i$ denotes the nearest neighbors of site $i$. As usual, one MCS refers to $N$ attempts at spin updates.

In the XY system, the in-plane rotation of the spins allows the formation of stable topological defects like vortices and anti-vortices. These defects are non-trivial zeros of the magnetization field \cite{puri2009kinetics,Bray1994}, with a core where the magnetization field vanishes. In $d=2$ the defect core is a point object, while in $d=3$ it extends to form vortex and anti-vortex strings. We identify these defects in a $d=2$ spin configuration by finding winding numbers $w$ on the square plaquettes of the lattice:
\be
\label{eq4}
w = \frac{1}{2 \pi} \oint_C \mathbf{\nabla} \theta \cdot \text{d}\mathbf{l},
\ee
where the contour $C$ is chosen counter-clockwise. The usual convention is that the plaquette has a vortex (or anti-vortex) if $w = +1$ (or $-1$). Notice that an anti-vortex is the counterpart of a vortex and has the same energetic cost. At high temperatures, unbound vortices and anti-vortices are free-energetically favorable. At low temperatures, these exist only as tightly bound vortex-anti-vortex pairs. When a vortex and anti-vortex come close, they annihilate and leave behind a local region of homogeneous magnetization \cite{PhysRevLett.97.177202}. Therefore, after a quench from high to low temperatures, the XY system evolves to the new equilibrium state via the annihilation of vortex and anti-vortex defects.

From the evolution morphology $\{\mathbf{S}_i(t)\}$ at a certain $t$, we determine the correlation function $C(\mathbf{r},t)$, defined as
\be
\label{eq5}
C(\mathbf{r} ,t) = \frac{1}{N} \sum_{\mathbf{R}} \left[\overline{\left<\mathbf{S}_{\mathbf{R}}(t) \cdot \mathbf{S}_{\mathbf{R+r}}(t)\right>}-\overline{\left<\mathbf{S}_{\mathbf{R}}(t)\right>} \cdot \overline{\left<\mathbf{S}_{\mathbf{R+r}}(t)\right>}\right],
\ee
where $\overline{\left< .... \right>}$ denotes an average over independent initial conditions and disorder realizations. As our system is homogeneous and isotropic, $C(\mathbf{r} ,t)$ can be spherically averaged to make it directionally independent, say $C(r,t)$. This quantity exhibits dynamical scaling if the system is characterized by a single length scale $L(t)$ \cite{puri2009kinetics}, i.e.,
\be
\label{cf_scale}
C(r,t) = f\left(\frac{r}{L(t)}\right) ,
\ee
where $f(x)$ is a scaling function. In experiments, one usually measures the structure factor $S(\mathbf{k} ,t)$, which is the Fourier transform of $C(\mathbf{r} ,t)$. The spherically-averaged structure-factor $S(k,t)$ has the scaling form: 
\be
\label{sf_scale}
S(k,t) = L(t)^d g\left(k L(t)\right) ,
\ee
where $g(p)$ is the scaling function.

Bray and Puri \cite{bray1991asymptotic}, and Toyoki \cite{toyoki1992structure} (BPT) have predicted the asymptotic form of $f(x)$ for a disorder-free system with $O(n)$ symmetry:
\be
\label{bpt}
f_{\rm BPT}(x) = \frac{n \gamma}{2 \pi} \left[B\left(\frac{n+1}{2},\frac{1}{2}\right)\right]^2 F\left(\frac{1}{2},\frac{1}{2};\frac{n+2}{2};\gamma^2\right) .
\ee
Here, $\gamma = \exp(-x^2)$, $B(a,b) = \Gamma(a) \Gamma(b)/ \Gamma(a+b)$ is the beta function, and $F(a,b;c;z)$ is the hypergeometric function. This scaling function for short distances $r\ll L$ contains a singular term of order $(r/L)^n$ for odd $n$, with an additional $\ln (r/L)$ factor for even $n$ \cite{bray1991asymptotic,toyoki1992structure}. This implies, through simple power counting, a power-law tail in the structure factor, i.e., $g(p) \sim p^{-(d+n)}$ for $kL\gg 1$. This is referred to as the {\it generalized Porod law} \cite{porod1982small,oono1988large}.

Another nonequilibrium quantity of interest is the {\it two-time autocorrelation function}, defined as
\be
\label{auto_eq}
A(t,t_w) = \frac{1}{N} \sum_{i=1}^{N} \left[\overline{\left<\mathbf{S}_i(t_w) \cdot \mathbf{S}_{i}(t)\right>}-\overline{\left<\mathbf{S}_i(t_w)\right>} \cdot \overline{\left<\mathbf{S}_i(t)\right>}\right],
\ee
where $t_w < t$ is the {\it waiting time}.  This quantity is crucial for studying the aging behavior \cite{zannetti2014aging}, i.e., how the relaxation processes slow down as the system ages. In general, the quantity $A(t,t_w)$ is known to  consist of two distinct parts, namely a {\it stationary part} $A_{\rm st}$ and an {\it aging part} $A_{\rm ag}$ \cite{zannetti2014aging,kumar2020growth}. The stationary part corresponds to equilibrium fluctuations, and exhibits {\it time-translational invariance} (TTI). Thus, it depends only on the time-difference $t-t_w$ and contributes when $t-t_w \ll t_w$. For large time separations ($t-t_w\gg t_w$), the system shows aging, which is accompanied by the loss of TTI and a breakdown of the {\it fluctuation-dissipation theorem} (FDT). The aging contribution has the scaling form \cite{zannetti2014aging,henkel2011non}:
\begin{equation}
\label{aging}
A_{\rm ag}(t,t_w) = h \left(\frac{L(t)}{L(t_w)}\right), 
\end{equation}
where $h(y)\sim y^{- \lambda_C}$ for $y \gg 1$ and the exponent $\lambda_C$ is known as the {\it autocorrelation exponent} \cite{fisher1988nonequilibrium}. 

Let us next discuss the framework for analyzing the asymptotic behavior of $L(t)$. We define $L(t)$ as the distance over which $C(r,t)$ decays to 0.2 of its maximum value, i.e., $C(L,t)=0.2 \times C(0,t)$. (For notational convenience, we write $L(t,\Delta)$ as $L(t)$, except when the discussion requires the $\Delta$-dependence.) Lippiello et al. \cite{lippiello2010scaling,corberi2011growth,corberi2012crossover} developed a novel method to study the crossover of $L(t,\Delta)$ from an early-time algebraic behavior $(L \sim t^{1/z})$ to the asymptotic behavior in the presence of disorder $\Delta$. In their formulation, 
\be
L(t, \Delta)= t^{1/z}\mathcal{F} ( \Delta t^{1/\varphi}) ,
\ee
where $\mathcal{F}(x)$ is the crossover function which dominates at long times. Thus,
\be
\mathcal{F}(x)\sim x^{-\varphi/z} \mathcal{\widetilde F}(x^\varphi), \quad x\to \infty ,
\ee
with the crossover exponent $\varphi >0$. In the inverted form, this scaling hypothesis for $L(t,\Delta)$ can be rewritten as 
\be
\label{invert}
t=L^z\mathcal{G}(L/\ell) ,
\ee
with the crossover length $\ell \sim \Delta^{-\varphi/z}$. The scaling function $\mathcal{G}(y)$ behaves as 
\ba
\label{g_scale}
\mathcal{G}(y) &= & \text{constant}, \quad y \ll 1, \nonumber \\
&=& y^{-z} \mathcal{\widetilde G} (y), \quad y \gg 1,
\ea
At $y \simeq 1$, i.e., $L(t, \Delta) \simeq \ell(\Delta)$, a crossover occurs from the early-time algebraic behavior to the asymptotic regime. The function $\mathcal{G}(y)$ is related to $\mathcal{F}(x)$ as $\mathcal{G}(y) = [\mathcal{F}(x)]^{-z}$, with  $\mathcal{\widetilde G} (y) =\mathcal{\widetilde F}^{-1} (y)$ being the inverse of $\mathcal{\widetilde F}$.

The significant quantity to predict the asymptotic growth is the {\it effective dynamic exponent}, defined as
\begin{equation}
\label{growth_exp}
z_{\rm eff} (t, \Delta)=\left [\frac{\partial \ln L (t, \Delta)}{\partial \ln t}\right]^{-1}. 
\end{equation}
The corresponding growth exponent is $\theta_{\rm eff} = 1/z_{\rm eff}$. Using Eq.~\eqref{invert}, $z_{\rm eff}$ can be expressed in terms of $y=L/\ell$ as 
\be
\label{eff_growth_L}
z_{\rm eff} (y) = \frac{\partial \ln t}{\partial \ln L}= z + \frac{\partial \ln \mathcal G(y)}{\partial \ln y}.
\ee
Then, using Eq.~\eqref{g_scale}, we obtain
\ba
\label{eff_growth}
z_{\rm eff}(y) &=& z, \quad y\ll 1, \nonumber \\
&=& \frac{\partial \ln \mathcal {\widetilde G} (y)}{\partial \ln y}, \quad y \gg 1 .
\ea
The condition $y \gg 1$ corresponds to the late-time behavior, which requires us to understand the function $\mathcal {\widetilde G} (y)$. Thus, we need to obtain ${\widetilde G}(y)$ from the numerical data to determine the asymptotic form of the growth law.

In the next section, we present numerical results from our simulations in $d=2,3$. The system sizes are $N=1024^2$ in $d=2$, and $128^3$ in $d=3$. The statistical data is averaged over 25 independent runs in $d=2$, each with different initial configurations of spins $\{\mathbf S_i(0)\}$ and random fields $\{\boldsymbol \epsilon_i \}$. The random field configuration is fixed for a run. In $d=3$, we perform 20 independent runs. We study the scaling behavior of all quantities introduced in this section.

\section{Detailed Numerical Results}
\label{simulations}

\subsection{d=2 RFXYM}
\label{2drfxym}

\begin{figure}[t]
\centering
\includegraphics[width=0.9\linewidth]{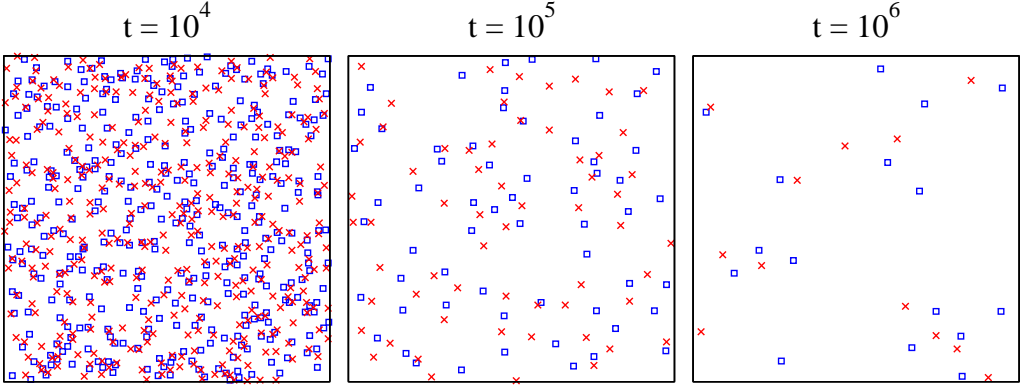}
\caption{Evolution snapshots of defects in the $d=2$ RFXYM with $\Delta=0.1$. The vortices and anti-vortices are labeled as squares (in blue) and crosses (in red), respectively. For a better view, we show only a $512^2$ corner of the full $1024^2$ system in each snapshot.}
\label{fig1}
\end{figure}

Let us start by presenting some evolution snapshots in terms of topological defects like vortices and anti-vortices. Fig.~\ref{fig1} shows the typical snapshots of vortex/anti-vortex evolution in the RFXYM for $\Delta=0.1$
with the evolution time $t$ in MCS. The defect density decreases with time, i.e., the system becomes more ordered via the annihilation of vortices and anti-vortices. However, unlike the Ising model where the magnetization is approximately homogeneous inside the domains ($+1$ or $-1$), the ``domains'' in the XY model are identified by the separation between different defect pairs. The spins are approximately parallel at the boundaries between defects, and they rotate strongly at defect cores (See the vector plots in Appendix~\ref{vector_plot}).

In Fig.~\ref{2d_fixt}, we show the $\{\theta_i\}$
\begin{figure}[t]
\centering
\includegraphics[width=0.9\linewidth]{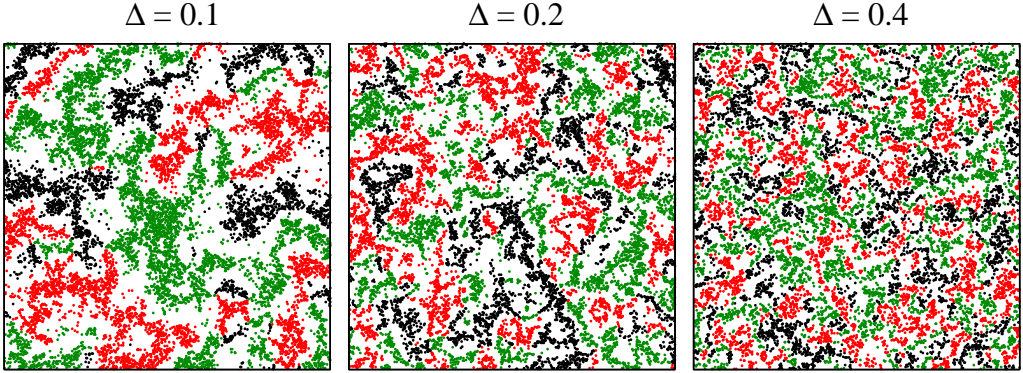}
\caption{Evolution snapshots of $\{\theta_i\}$ at $t = 10^6$ MCS in the $d=2$ RFXYM with $\Delta = 0.1,0.2,0.4$. The regions with angle $\theta_i$ are colored according to the interval $[\theta_0-0.1,\theta_0+0.1]$ in which they lie. We use the following color coding: $\theta_0=\pi/3$ (black), $\theta_0=\pi$ (red/gray), $\theta_0=5\pi/3$ (green/light gray). In each snapshot, we show only a $512^2$ corner of the full $1024^2$ system size.}
\label{2d_fixt}
\end{figure}
configurations at a fixed $t=10^6$, and for different $\Delta$: 0.1, 0.2 and 0.4. Each $\theta_i \in (0,2\pi)$ and is colored according to three different windows of equal size. These small windows are specified in the figure caption. A junction point of the three colors corresponds to a vortex or anti-vortex, depending on the direction of rotation. The defect density increases with disorder amplitude, which signals a slowing down of the domain growth process at higher $\Delta$.

\begin{figure}[t]
	\centering
	\includegraphics[width=1.0\textwidth]{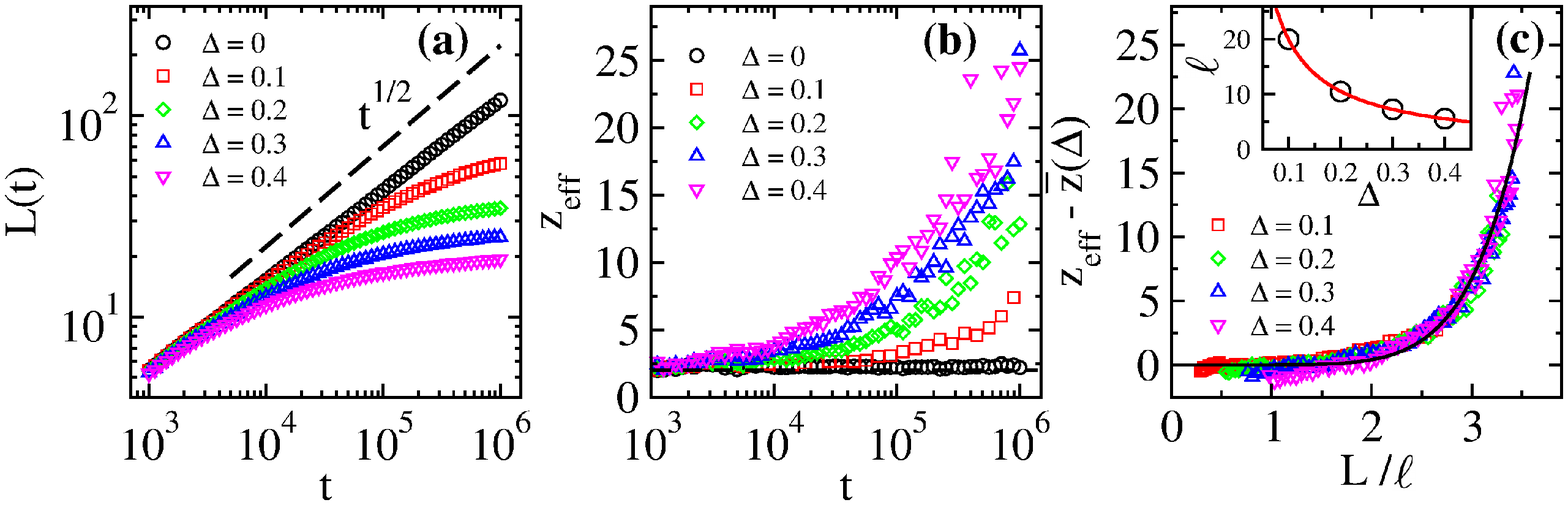}
	\caption{(a) Plot of $L(t)$ vs. $t$ on a log-log scale for the $d=2$ RFXYM at various values of $\Delta$, as specified. The dashed line of slope 0.5 denotes the LAC growth law. Notice that the $\Delta = 0$ data has an effective growth exponent less than 1/2, due to a logarithmic correction. (b) Plot of effective dynamic exponent $z_{\rm eff} = [\partial \ln L/  \partial \ln t]^{-1}$ vs. $t$ for the data shown in (a). The dashed line denotes $\bar z=2$. (c) Scaling of $z_{\rm eff} - \bar{z}(\Delta)$ vs. $L/ \ell (\Delta)$, where $\ell (\Delta)$ is chosen to enable a data collapse. The solid line in the main frame is the best power-law fit to the scaled data, i.e., $z_{\rm eff} - \bar{z} \simeq 0.0034 (L/\ell)^{6.8 \pm 0.3}$. The inset shows $\ell$ vs. $\Delta$ with the best fit $\ell \sim \Delta^{-0.92}$.}
	\label{Lt_2d}
\end{figure}

We are primarily interested in the growth law, i.e., the behavior of $L(t)$ as a function of $t$. In Fig.~\ref{Lt_2d}(a), we plot $L(t)$ vs. $t$ on a log-log scale. we consider only data for $t > 10^3$ so as to eliminate the transients associated with domain formation. For the pure XY model ($\Delta = 0$), the
growth law exhibits a logarithmic correction in $d=2$: $L(t) \sim (t/\ln t)^{1/2}$ \cite{yurke1993coarsening}. This correction reduces the effective exponent to a value lower than 1/2, which is the usual growth law for nonconserved kinetics. For $\Delta > 0$, $L(t)$ initially grows in a power-law fashion, but slows down at late times. To study this behavior, we determine the effective dynamic exponent $z_{\rm eff}$, as defined in Eq.~\eqref{growth_exp}.

In Fig.~\ref{Lt_2d}(b), we plot $z_{\rm eff}$ vs. $t$. For $\Delta=0$, $z_{\rm eff}$ shows a flat behavior at all $t$ around the expected value, i.e., $z_{\rm eff}(\tau,\Delta=0) \simeq 2$. For $\Delta>0$, $z_{\rm eff}$ is flat in an intermediate regime of time at a $\Delta$-dependent value, say $\bar z (\Delta)$ \cite{Paul_2004,ppr05}, which is followed by an upward trend in $z_{\rm eff}$ at late times. To identify the pre-asymptotic exponent $\bar z (\Delta)$, we fit a power-law to the initial points of the data sets in Fig.~\ref{Lt_2d}(a). The time window in which $z_{\rm eff}$ is flat becomes smaller as $\Delta$ increases. For small $\Delta$, $z_{\rm eff}$ is flat over nearly one decade of time $[10^3, 10^4]$. The values of $\bar z (\Delta)$ with error bars are specified in Table~\ref{exp_tab}. This upward trend corresponds to a slowing down in growth as the growth exponent $\theta_{\rm eff} = 1/z_{\rm eff}$. As discussed by Lippiello et al. \cite{lippiello2010scaling,corberi2011growth,corberi2012crossover}, this signals the onset of an asymptotic logarithmic growth regime. To identify this crossover behavior using Eq.~\eqref{eff_growth_L}, we plot $z_{\rm eff} - \bar{z}(\Delta)$ as a function of $L/ \ell (\Delta)$ in Fig.~\ref{Lt_2d}(c). If Eq.~\eqref{eff_growth_L} holds, an appropriate choice of $\ell (\Delta)$ should collapse the data sets for different $\Delta$. Thus, for each $\Delta$, we determine the crossover length $\ell(\Delta)$ by requiring that the data collapses onto a scaling function, as shown in Fig.~\ref{Lt_2d}(c). The $\Delta$-dependence of $\ell$ is plotted in an inset of this figure, and is fitted well by $\ell  \sim \Delta^{-0.92}$ as shown by the solid curve. We fit the scaled data in Fig.~\ref{Lt_2d}(c) to a power-law: $z_{\text{eff}} - \bar{z} = a y^{\psi}$, where $y = L/\ell$. This
yields $a \simeq 0.0034$ and the exponent $\psi \simeq 6.8 \pm 0.3$. Then, from Eq.~\eqref{eff_growth_L}, it follows that
\begin{equation}
\frac{\partial \ln \mathcal G(y)}{\partial \ln y}=a y^{\psi} ~ \Rightarrow ~ {\mathcal G}(y)\sim \exp \left(\frac{a}{\psi} y^{\psi} \right). 
\label{exp_g}
\end{equation}

With this functional form of ${\mathcal G}(y)$, it is easy to derive the asymptotic form of the growth law using Eqs.~\eqref{invert}-\eqref{g_scale}, i.e.,
\begin{equation}
L(t,\Delta)\sim \Delta^{-0.92} \left[\frac{\psi}{a} \ln \left(t \Delta^{0.92\bar z}\right) \right]^{1/\psi}.
\end{equation}
This demonstrates that the  growth law crosses from the pre-asymptotic algebraic behavior $L(t,\Delta) \sim t^{1/\bar z(\Delta)}$ to the asymptotic logarithmic behavior $L(t,\Delta) \sim (\ln t)^{1/\psi}$ with the logarithmic growth exponent $\psi=6.8 \pm 0.3$. The crossover takes place at $L(t,\Delta) \simeq \ell(\Delta)$, where $\ell \sim \Delta^{-0.92}$, i.e., for large disorder amplitudes the crossover occurs at smaller length scales. We had also noticed this in Fig.~\ref{Lt_2d}. This behavior of the growth law is consistent with our earlier studies of disordered Ising models \cite{lippiello2010scaling,corberi2011growth,corberi2012crossover}. Of course, the value of the logarithmic growth exponent $1/\psi$ depends on the specific system being considered.

\begin{table}[b]
\caption{Values of exponents $B$, $\lambda_C$, and $\bar{z}(\Delta)$ for the RFXYM in $d=2,3$. The exponents $\lambda_C$ and corresponding error bars are estimated from power-law fits to $A(t,t_w)$ in the scaling regime $[L(t) \gg L(t_w)]$. For these fits, we use the smallest value of $t_w$ to obtain the largest range of the scaling variable. To estimate $\bar{z}(\Delta)$ and their error bars, we perform fits in the pre-asymptotic regime of the growth law: $L(t,\Delta)\sim t^{1/\bar{z}(\Delta)}$.}
	\label{exp_tab}
	\begin{center}
		\begin{tabular}{| c | c | c | c | c | c |  c |} 
			\hline 
			\multicolumn{4}{|c|}{$d=2$}
			& \multicolumn{3}{|c|}{$d=3$}
			\\ \hline
			$\Delta$ & $B$ & $\lambda_C$ & $\bar{z}$ & $\Delta$ & $\lambda_C$ & $\bar{z}$ \\
			\hline \hline
			0 & 0.037 & 1.20 $\pm$ 0.01 & 2.22 $\pm$ 0.01 & 0 & 1.71 $\pm$ 0.03 & 2.11 $\pm$ 0.05 \\
			\hline
			0.1 & 0.651 & 1.33 $\pm$ 0.02 & 2.40 $\pm$ 0.03 & 0.5 & 1.45 $\pm$ 0.02 & 2.26 $\pm$ 0.04  \\
			\hline
			0.2 & 0.528 & 1.31 $\pm$ 0.01 & 2.63 $\pm$ 0.07 & 1 & 1.05 $\pm$ 0.03 & 2.49 $\pm$ 0.09 \\
			\hline
			0.3 & 0.250 & 1.28 $\pm$ 0.02 & 2.89 $\pm$ 0.04 & 1.5 & 0.77 $\pm$ 0.03 & 2.74 $\pm$ 0.06 \\
			\hline
			0.4 & 0.150 & 1.16 $\pm$ 0.02 & 3.38 $\pm$ 0.05 & 2 & 0.61 $\pm$ 0.02 & 3.20 $\pm$ 0.16 \\
			\hline
		\end{tabular}
	\end{center}
\end{table}

\begin{figure}[t]
\centering
\includegraphics[width=0.8\linewidth]{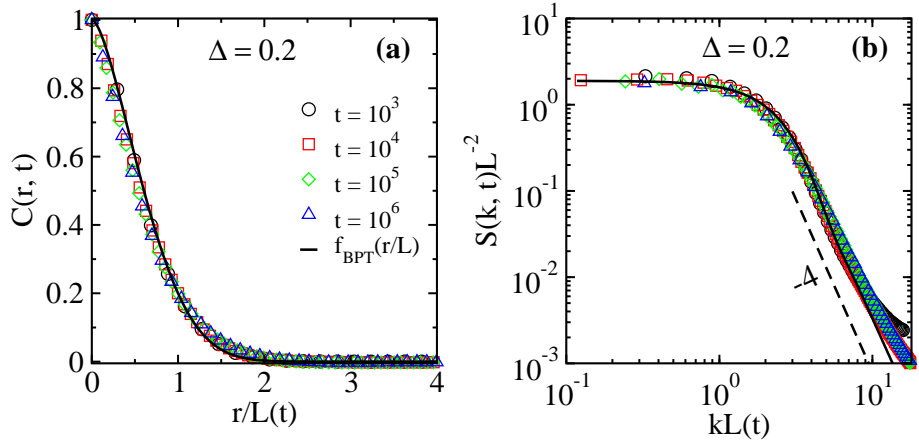}
\caption{Scaling plots of the (a) correlation function $C(r,t)$ vs. $r/ L(t)$; and (b) structure factor $S(k,t) L(t)^{-2}$ vs. $kL(t)$ for the $d=2$ RFXYM with $\Delta = 0.2$. We superpose data for distinct times, as specified. The solid curve in (a) denotes the BPT function $f_{\rm BPT} (r/L)$ for $n=2$, as defined in Eq.~\eqref{bpt}. The solid line in (b) is its Fourier transform in $d=2$.  The dashed line in (b) denotes the generalized Porod law: $g(p) \sim p^{-(d+n)}$ for $d=n=2$.}
\label{corr_2d}
\end{figure}

\begin{figure}[t]
\centering
\includegraphics[width=0.8\linewidth]{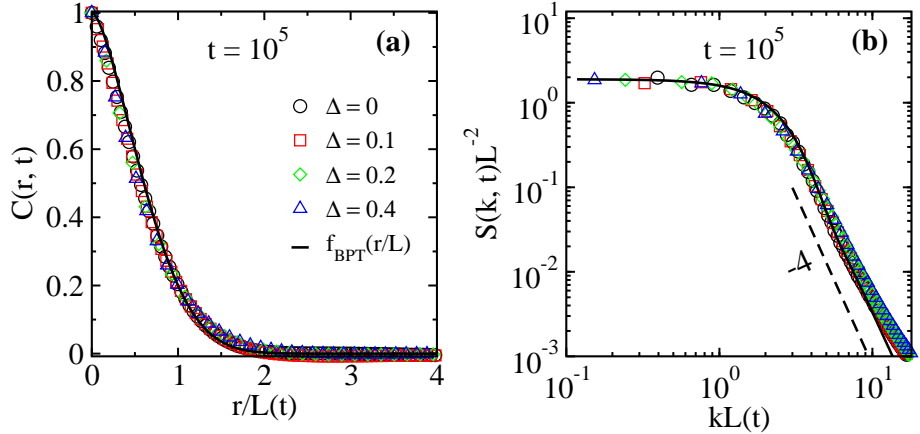}
\caption{Analogous to Fig.~\ref{corr_2d}, but scaled data for (a) $C(r,t)$, and (b) $S(k,t)$ is plotted for different $\Delta$ at $t=10^5$.}
\label{corr_dis_2d}
\end{figure}

We further analyze the scaling functions introduced in Sec.~\ref{method}. In Fig.~\ref{corr_2d}, we plot the dynamical scaling forms of (a) the correlation function $C(r,t)$ vs.
$r/L(t)$; and (b) the structure factor $S(k,t) L(t)^{-2}$ vs. $kL(t)$ for $\Delta=0.2$. An excellent data collapse in both panels confirms that the system is characterized by a single length scale $L(t)$. The solid curve in (a) denotes the BPT function $f_{\rm BPT} (r/L)$, as defined in Eq.~(\ref{bpt}), for $n=2$. The solid curve in (b) is the Fourier transform of $f_{\rm BPT}$ in $d=2$ ($\mathbf{p} = \mathbf{k}L$):
\begin{eqnarray}
g_{\rm BPT}(p)&=&\int \! d \textbf{x} \, e^{i \textbf{p} \cdot \textbf{x}} f_{\rm BPT}(x) \nonumber \\
&=& 2\pi \int_0^\infty dx~x f_{\rm BPT}(x) J_0(px) .
\label{ft}
\end{eqnarray}
Here,
\be
J_0(s) = \frac{1}{\pi} \int_0^\pi~d\theta \cos (s\cos \theta)
\ee
is the Bessel function of the first kind of order 0.  Clearly, $f_{\rm BPT}$ in (a), and $g_{\rm BPT}$ in (b), describes the numerical data very well. (Recall that the BPT result was obtained in the context of a disorder-free system. This suggests that $f(x)$ and $g(p)$ are independent of disorder.) The dashed line in (b) denotes the generalized Porod law behavior of $g(p)$ at large $p$, i.e., $g(p) \sim p^{-(d+n)}$ for $d=n=2$. Let us next examine the universality of these scaling functions with respect to $\Delta$. Fig.~\ref{corr_dis_2d} shows the plots of $f(r/L)$ and $g(kL)$ at a fixed value of
$t=10^5$ but for different $\Delta$ values. A neat data collapse confirms universality in the scaling functions for $C(r,t)$ and $S(k.t)$.

\begin{figure}[t]
	\centering
	\includegraphics[width=0.75\linewidth]{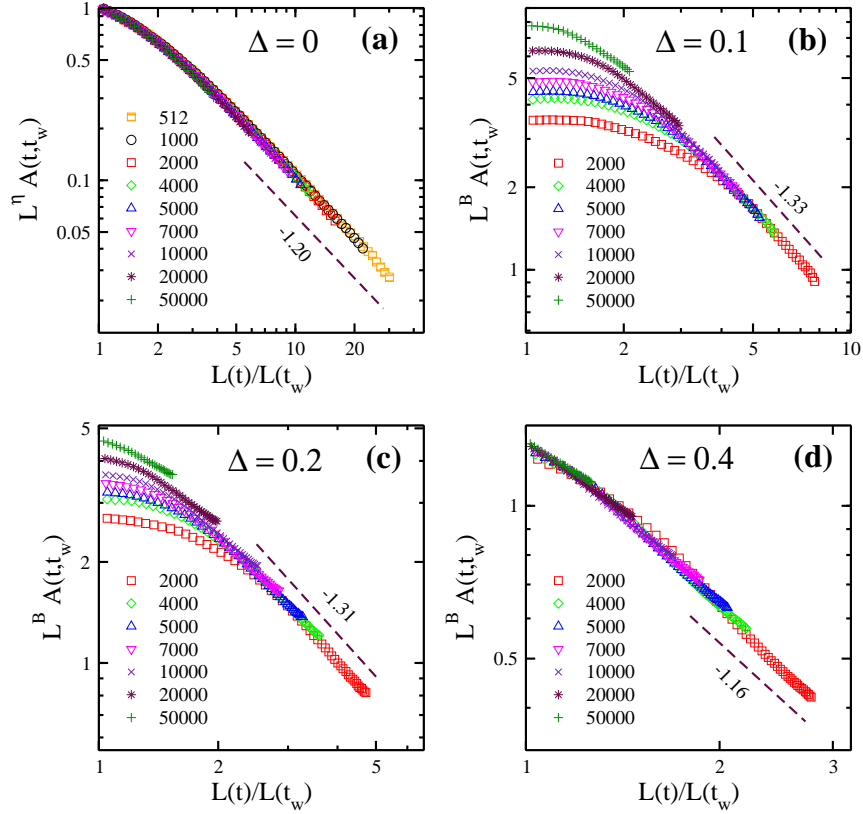}
	\caption{(a) Scaling plot of autocorrelation function $A(t,t_w)$ for $\Delta=0$, using the scaling form in Eq.~\eqref{autoscale_L}. We plot $L(t)^\eta A(t,t_w)$ vs. $L(t)/L(t_w)$ for different $t_w$, as specified. The value of $\eta(T) \simeq 0.037$ \cite{berche2003bulk}. (b)-(d) Scaling plot of $A(t,t_w)$ for $\Delta \neq 0$. The analog of exponent $\eta$ in this case is termed the aging exponent $B$. The values of $B$ are adjusted to enable a data collapse at large $L(t)/L(t_w)$. They are specified in Table~\ref{exp_tab}. The dashed lines indicate a power-law decay at large $t$, i.e., $A_{\rm ag}(t,t_w) \sim [L(t)/L(t_w)]^{-\lambda_C}$, where $\lambda_C$ is mentioned in the frames.}
	\label{autoscale_2d}
\end{figure}

To check if this universality extends to the two-time observables, we inspect the scaling properties of
the autocorrelation function $A(t,t_w)$, as defined in Eq.~(\ref{auto_eq}). Let us first focus on the pure case with $\Delta=0$. Regarding the stationary contribution to $A(t,t_w)$, the pure XY system exhibits a quasi-equilibrium regime, which is well represented by a power-law \cite{berthier2001nonequilibrium}:
\be
A_{\rm st}(t,t_w) \sim (t-t_w)^{-\eta (T)/z}, \quad t-t_w \ll t_w .
\ee
Here, $\eta (T)$ is the exponent of the equilibrium spin-spin correlation function and $z=2$ is the dynamical exponent. At our quench temperature $T=0.2$, $\eta \simeq 0.037$ \cite{berche2003bulk}. The aging contribution is a function of the variable $L(t)/L(t_w)$. With these two contributions, the complete form of $A(t,t_w)$ has been conjectured to obey so-called {\it non-simple} scaling \cite{berthier2001nonequilibrium,abriet2004off}:
\begin{equation}
A(t,t_w)\sim (t-t_w)^{-\eta (T)/z} h[L(t)/L(t_w)] .
\label{autoscale_pure}
\end{equation}
At long times $(t \gg t_w)$, Eq.~(\ref{autoscale_pure}) becomes
\begin{equation}
A(t,t_w)\sim L(t)^{-\eta} h[L(t)/L(t_w)].
\label{autoscale_L}
\end{equation}

In Fig.~\ref{autoscale_2d}(a), we plot $L(t)^{\eta (T)} A(t,t_w)$ vs. $L(t)/L(t_w)$ for $\Delta=0$ and different $t_w$-values. The data collapse is excellent, thus confirming the scaling prediction in Eq.~(\ref{autoscale_L}). The dashed line in this figure gives the asymptotic power-law behavior of the scaling function: $h(y) \sim y^{- \lambda_C}$ for $y \gg 1$. The slope is estimated by using the data set for the smallest value of $t_w$ in (a) -- this yields the largest range of values for $L(t)/L(t_w)$. We find $\lambda_C \simeq 1.20$, which is comparable to the value $\lambda_C \simeq 1.086$ obtained in previous studies of the XY model \cite{berthier2001nonequilibrium,abriet2004off}. We test for the same scaling form for the disordered cases, as shown in Figs.~\ref{autoscale_2d}(b)-(d). In the disordered case, $\eta$ is treated as an adjustable parameter, which we term as the {\it aging exponent} $B$. With disorder, the relaxation of the system slows down and the data does not scale at short times. As in the pure case, the aging effects dominate at long times. In panels (b)-(d), we superimpose the data of $A(t,t_w)$ for different $t_w$-values by plotting $L(t)^B A(t,t_w)$ vs. $L(t)/L(t_w)$ with a suitable value of exponent $B$ that produces a scaling collapse in the aging regime. As expected, the aging function decays algebraically in the asymptotic limit, with the  autocorrelation exponent $\lambda_C$ depicted by a dashed line in each panel. The $\lambda_C$-exponents for different $\Delta$-values are consistent with the inequalities $d \geq \lambda_C \geq d/2$ \cite{fisher1988nonequilibrium,huse1989remanent,yeung1996bounds}. The estimated values of exponents $\lambda_C$ and $B$ with error bars are presented in Table~\ref{exp_tab}.

Moreover, Fig.~\ref{autoscale_2d} shows that the tail of the aging function depends explicitly upon the disorder amplitude. This demonstrates the breakdown of {\it superuniversality} (SU) \cite{pcp91,pp92} in the ordering dynamics of the RFXYM. (For SU to hold, we would expect all disorder effects to be captured by the length scale $L(t,\Delta)$. Thus, $f(x), g(p), h(y)$ etc. should all be independent of $\Delta$.) This is consistent with our earlier study on the RBXYM \cite{kumar2017ordering}.

\subsection{d=3 RFXYM}

Let us next present numerical results for the RFXYM in $d=3$. For reference purposes, we recall results for the XY model in $d=3$. The model undergoes a transition from the disordered state to a state with LRO at $T_c \simeq 2.202$ \cite{hasenbusch1990critical}. After a rapid quench to $T < T_c$, the growth process is characterized by the usual diffusive law $L(t) \sim t^{1/2}$ -- the logarithmic correction only arises in $d=2$.

\begin{figure}[t]
\centering
\includegraphics[width=0.95\textwidth]{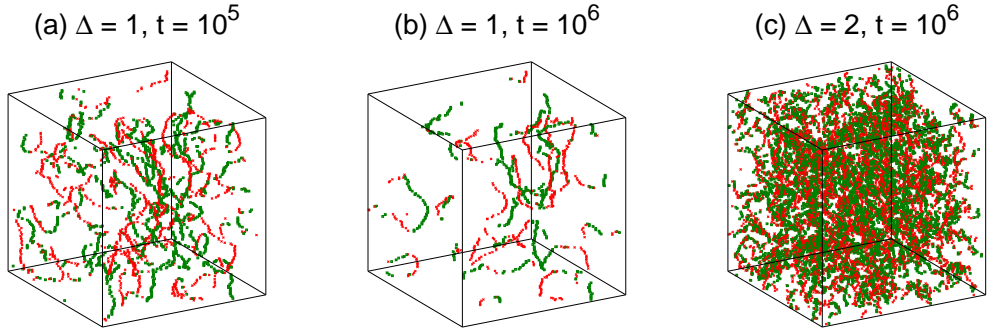}
\caption{Evolution snapshots of defects (vortex and anti-vortex strings) in the $d=3$ RFXYM. The system size is $128^3$. We show snapshots for (a) $\Delta=1,~t=10^5$; (b) $\Delta=1,~t=10^6$; and (c) $\Delta=2,~t=10^6$. The vortex strings and anti-vortex strings are marked red (crosses) and green (squares), respectively.}
\label{fig3d}
\end{figure}

For domain growth in the RFXYM, we first show evolution snapshots of the topological defects, which are strings \cite{mg90,mg92}. Fig.~\ref{fig3d} shows typical string configurations for (a) $\Delta=1,~t=10^5$; (b) $\Delta=1,~t=10^6$; and (c) $\Delta=2,~t=10^6$. Panels (a) and (b) show the evolution
for $\Delta=1$, whereas panels (b) and (c) show that the defect density increases on raising $\Delta$. The defects are trapped by the disorder sites, and the energy barriers are higher for larger disorder amplitudes.

\begin{figure}[t]
	\centering
	\includegraphics[width=1.0\textwidth]{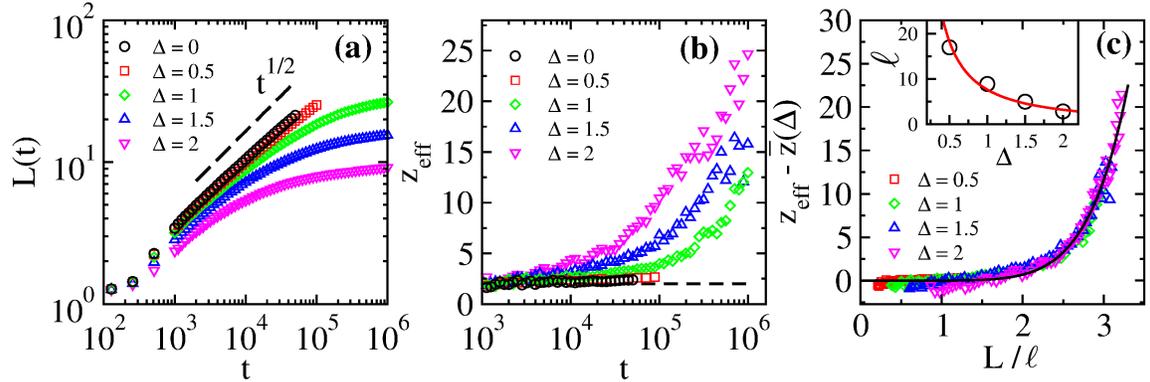}
	\caption{(a) Plot of $L(t)$ vs. $t$ on a log-log scale for the $d=3$ RFXYM with $\Delta$ as specified. The dashed line denotes the LAC growth law for the pure XY model. (b) Plot of effective dynamic exponent $z_{\rm eff} = [\partial \ln L/  \partial \ln t]^{-1}$ vs. $t$ for the data shown in (a). The dashed line denotes $\bar z=2$. (c) Scaling of $z_{\rm eff} - \bar{z}(\Delta)$ vs. $L/ \ell$ for different $\Delta$. The solid curve is the best power-law fit: $z_{\rm eff} - \bar{z} \simeq 0.0055 (L/\ell)^{6.9 \pm 0.2}$. The inset shows $\ell$ vs. $\Delta$ with the fit $\ell \sim \Delta^{-1.25}$.}
	\label{Lt_3d}
\end{figure}

Let us first investigate the domain growth law. In Fig.~\ref{Lt_3d}(a), we plot $L(t)$ vs. $t$ on a log-log
scale for various values of $\Delta$. We see that $L(t)$ for $\Delta =0$ follows the expected algebraic behavior $L(t) \sim t^{1/2}$. For nonzero $\Delta$, $L(t)$ initially grows algebraically ($L \sim t^{1/\bar z(\Delta)}$), and then becomes slower at late times. To understand the nature of the slower growth, we again compute the effective growth exponents $z_{\rm eff}$, as defined in Eq.~\eqref{growth_exp}. In Fig.~\ref{Lt_3d}(b), we plot these as a function of $t$ for $\Delta=0,0.5,1,1.5,2$. For $\Delta=0$, a dashed line is drawn corresponding to the value $\bar z=2$. For $\Delta \ne 0$, as in the $d=2$ case, $z_{\rm eff}$ is initially flat at $\bar z(\Delta)$, and then curves upward at late times. The values of $\bar z (\Delta)$ are specified in Table~\ref{exp_tab}.

In Fig.~\ref{Lt_3d}(c), we plot the subtracted exponents $z_{\rm eff}-\bar z (\Delta)$ as a function of $L/\ell$, with $\ell$ being adjusted to achieve a scaling collapse for different $\Delta$. The variation of $\ell$ with $\Delta$ is shown in the inset, and obeys $\ell \sim \Delta^{-1.25}$. The solid curve in the main panel of Fig.~\ref{Lt_3d}(c) is a power-law fit to the scaled data, i.e., $z_{\rm eff} - \bar{z} = ay^\psi \simeq 0.0055 y^{6.9 \pm 0.2}$ with $y=L/\ell$. This yields an exponential behavior for the scaling function $\mathcal{G}(y)$, as in Eq.~\eqref{exp_g}. Then, the asymptotic growth law is
\begin{equation}
L(t,\Delta)\sim \Delta^{-1.25} \left[\frac{\psi}{a} \ln \left(t\Delta^{1.25\bar z} \right)\right]^{1/\psi},
\end{equation}
where $a \simeq 0.0055$ and the logarithmic exponent $\psi \simeq 6.9 \pm 0.2$. The crossover length-scale from power-law behavior at early times to an asymptotic logarithmic growth is given by $L(t,\Delta) \simeq \ell(\Delta)$, with $\ell \sim \Delta^{-1.25}$.

\begin{figure}[h!]
\centering
\includegraphics[width=0.6\linewidth]{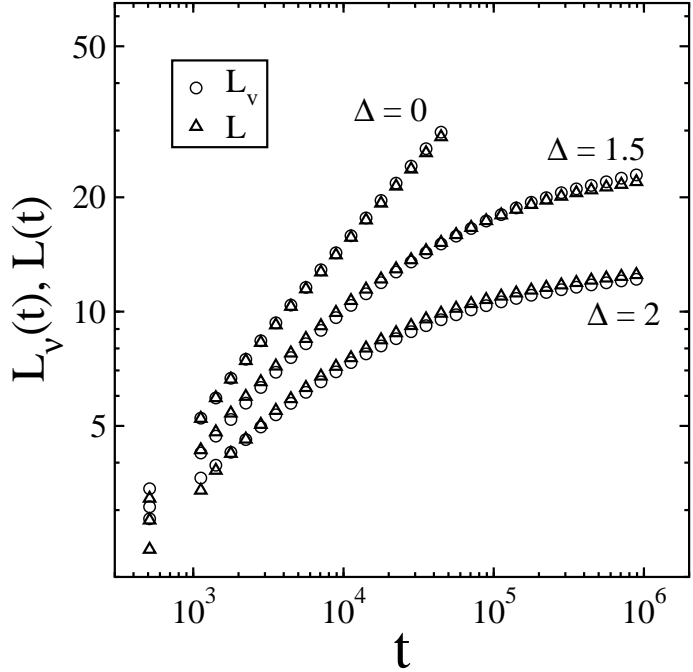}
\caption{Plot of $L_v(t)$ and $L(t)$ vs. $t$ on a log-log scale for the $d=3$ RFXYM with different $\Delta$. The data for $L(t)$ has been multiplied by a factor $\simeq 1.4$.}
\label{Lv_def}
\end{figure}

The inverse of the defect density is also proportional to the length scale, as seen in Figs.~\ref{fig1} and \ref{fig3d}. Therefore, it is relevant to analyze the length scale obtained from the defect density, say $L_v(t)$, and to compare it with $L(t)$ obtained from the decay of $C(r,t)$. If the vortex/anti-vortex defect density is $\rho_{\rm def}(t)$, then the average defect separation $L_v(t)$ is \cite{PhysRevE.49.4925}
\be
\label{Lv}
[L_v(t)]^n \sim \frac{1}{\rho_{\rm def}(t)},
\ee
where $n$ is the number of components of a vector spin ($n=2$ for the XY model). In Fig.~\ref{Lv_def}, we compare $L_v(t)$ from Eq.~(\ref{Lv}) and $L(t)$ for different $\Delta$ in $d=3$. The data sets for $L(t)$ are scaled by a factor $\simeq 1.4$ to account for the possibility of the two length scales having different prefactors. The data sets are seen to be in excellent numerical agreement.

Consequently, we see that the asymptotic nature of the growth law in both $d=2,3$ is consistent with logarithmic behavior. As a useful remark, let us compare the logarithmic growth for the RFXYM to that for the $d=3$ RBXYM \cite{kumar2017ordering}, where $\psi \simeq 1.16$. Therefore, the logarithmic growth in the RFXYM is clearly much slower than that for the RBXYM. However, we should stress that the RBXYM considered in Ref.~\cite{kumar2017ordering} had random exchange interactions which were always ferromagnetic. Therefore, that system has a low-temperature ferromagnetic phase with LRO. In the present case, we do not have LRO (and perhaps not even QLRO) at low temperatures.

\begin{figure}[t]
\centering
\includegraphics[width=0.8\linewidth]{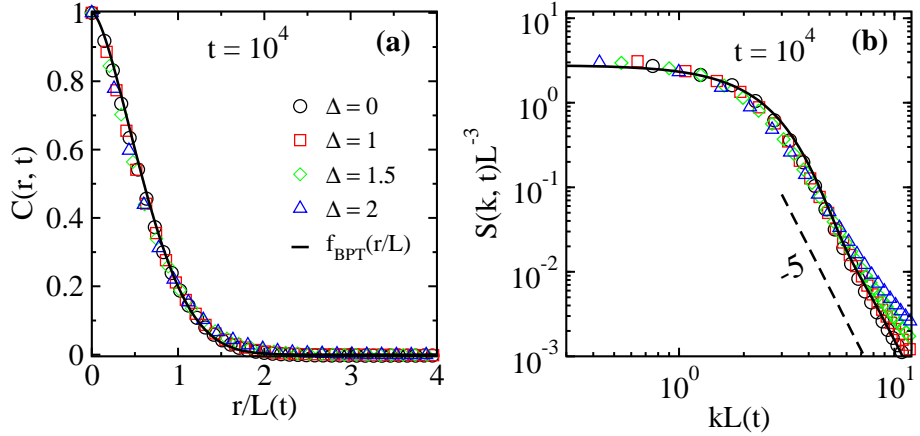}
\caption{Scaling plots of (a) $C(r,t)$ vs. $r/L(t)$; and (b) $S(k,t) L^{-3}$ vs. $kL(t)$ for the $d=3$ RFXYM with different $\Delta$ at $t=10^4$. The solid curve in (a) denotes the BPT function $f_{\rm BPT} (r/L)$ for $n=2$. The solid line in (b) is its Fourier transform in $d=3$. The dashed line in (b) denotes the generalized Porod law: $g(p) \sim p^{-(d+n)}$ for $d=3$ and $n=2$.}
\label{corr_dis_3d}
\end{figure}

We now come to the scaling functions. We have confirmed (not shown here) that $C(r,t)$ and $S(k,t)$ exhibit dynamical scaling for different values of $\Delta$. In Fig.~\ref{corr_dis_3d}, we plot the scaling functions $f(r/L)$ and $g(kL)$ at different $\Delta$
and a fixed $t=10^4$. The scaling functions for different $\Delta$ also collapse, as seen in Figs.~\ref{corr_dis_3d}(a)-(b). The solid curve in (a) is $f_{\rm BPT} (r/L)$, i.e., Eq.~\eqref{bpt} for $n=2$. The solid curve in (b) is its Fourier transform in $d=3$:
\begin{equation}
g_{\rm BPT}(p)= 4\pi \int_0^\infty~dx~x^2 f_{\rm BPT}(x) j_0(px) ,  
\end{equation}
where
\be
j_0(u)= \frac{\sin u}{u}
\ee
is the zeroth-order spherical Bessel function. Both $f_{\rm BPT}$ and $g_{\rm BPT}$ agree very well with the numerical data for $f(r/L)$ and $g(kL)$, respectively. The scaled data for $S(k,t)$ shows the generalized Porod tail: $g(p) \sim p^{-(d+n)} \sim p^{-5}$ for $d=3, n=2$.

\begin{figure}
\centering
\includegraphics[width=0.7\linewidth]{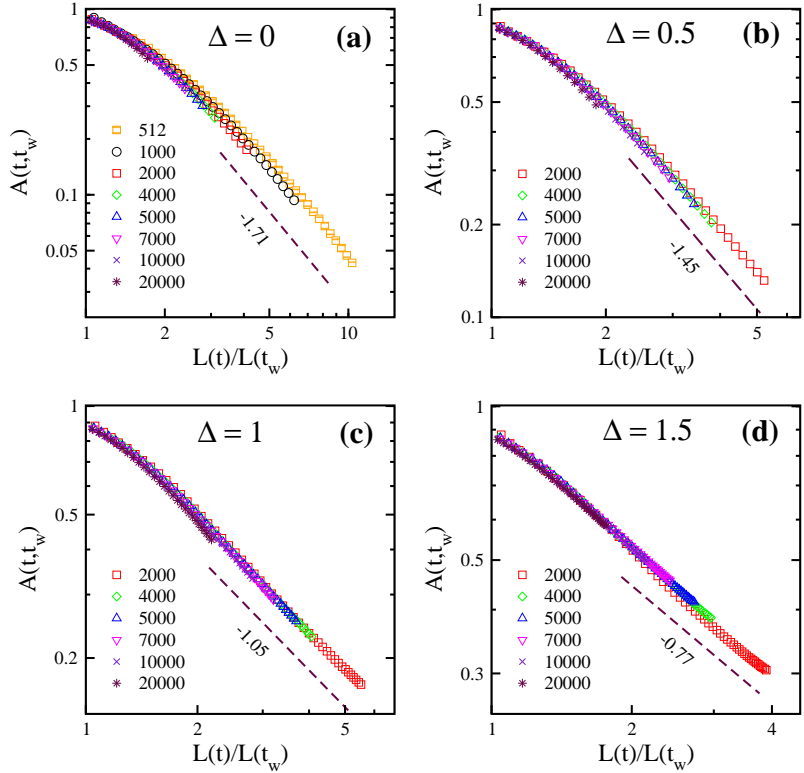}
\caption{Scaling plots of $A(t,t_w)$ vs. $L(t)/L(t_w)$ for the $d=3$ RFXYM with (a) $\Delta=0$, (b) $\Delta=0.5$, (c) $\Delta=1$, and (d) $\Delta=1.5$. The dashed lines denote the asymptotic behavior of the aging function, i.e., $A_{\rm ag}(t) \sim [L(t)/L(t_w)]^{-\lambda_C}$. The values of $\lambda_C$ are specified in the frames and in Table~\ref{exp_tab}.}
\label{autocorr_3d}
\end{figure}

To check for SU, we next examine the scaling of the autocorrelation function $A(t,t_w)$. As in the $d=2$ case, let us first look at the disorder-free system with $\Delta = 0$. In this case, $A(t,t_w)$ takes the following {\it simple} scaling form \cite{abriet2004off3d}:
\begin{equation}
A(t,t_w)=m_{\rm eq}^2(T) h[L(t)/L(t_w)] .
\label{auto_3d}
\end{equation}
Here, $m_{\rm eq}^2(T)$ is the square of the equilibrium magnetization at temperature $T$, which is the stationary contribution to $A(t,t_w)$ for short times $t-t_w \ll t_w$. [Notice that $\lim \limits_{t \rightarrow t_w} A(t,t_w) = m_{\rm eq}^2(T)$, whereas $A(t_w,t_w)=1$ from the definition in Eq.~\eqref{auto_eq}.] Fig.~\ref{autocorr_3d}(a) illustrates
the scaling of $A(t,t_w)$ for $\Delta=0$, according to the form in Eq.~\eqref{auto_3d}. Using the same form with $m_{\rm eq}^2(T)$ replaced by a parameter $C$, panels (b)-(d) are the plots of $A(t,t_w)$ as a function of $L(t)/L(t_w)$ for the disordered cases. The prefactor $C \simeq 0.90$ in all panels (a)-(d). A clean data collapse for different $t_w$-values can be seen in all the panels. For large times, the system enters the aging regime. The asymptotic power-law behavior in this regime is denoted by a dashed line in each plot. We estimate the slope by using the data set for the smallest value of $t_w$ in (a)-(d). The slope for each $\Delta$ is stated in the respective plot, and measures the autocorrelation exponent $\lambda_C$. The values of $\lambda_C$ (with error bars) are listed in Table~\ref{exp_tab}. The estimate of $\lambda_C \simeq 1.71 \pm 0.03$ for $\Delta=0$ is consistent with the value $1.7 \pm 0.1$ quoted in Ref.~\cite{abriet2004off3d}.   

An interesting point to note here is that $\lambda_C$ for $\Delta \neq 0$ violates the lower bound $\lambda_C \geq d/2$ \cite{yeung1996bounds}. A similar violation has also been observed numerically in the randomly diluted $d=2$ Ising model \cite{paul2007superaging}. This feature is at variance with the $d=2$ RFXYM discussed in Sec.~\ref{2drfxym}, where the inequality $\lambda_C \geq d/2$ is found to be obeyed -- see Table~\ref{exp_tab}. As the aging function for different $\Delta$ decays algebraically with distinct slopes $\lambda_C$, this rules out the possibility of SU.

\section{Summary and Discussion}
\label{summary}

We now conclude this paper by summarizing our results, and presenting a perspective for future work. The presence of disorder is inevitable in experimental systems. In this context, spin models with various types of quenched disorder have received considerable theoretical attention. Among these, the RFXYM is a model with several experimental realizations. In this paper, we undertake the first study (to the best of our knowledge) of phase ordering kinetics in the $d=2,3$ RFXYM. For this purpose, we performed comprehensive MC simulations. In our simulations, an initially disordered system (at $T = \infty$) undergoes a deep quench at time $t=0$. The system evolves up to $t=10^6$ MCS via the Metropolis algorithm with nonconserved kinetics. The evolution is characterized by the annihilation of topological defects of dimension $d-2$, i.e., vortices in $d=2$ and vortex strings in $d=3$. In the presence of quenched disorder of amplitude $\Delta$, the evolution slows down due to the trapping of coarsening defects by disorder-induced energy barriers.

We examined the asymptotic form of the growth law $L(t,\Delta)$ of characteristic length scale in the system. In both $d=2,3$, $L(t,\Delta)$ shows a crossover from an algebraic behavior with a $\Delta$-dependent exponent $(L \sim t^{1/\bar z (\Delta)})$ to a logarithmic behavior $L \sim (\ln t)^{1/\psi}$. The exponent $\psi \simeq 6.8$ in $d=2$, and $\psi \simeq 6.9$ in $d=3$.

We characterized the evolution morphology $\{\mathbf{S}_i(t)\}$ by the spatial correlation function $C(r,t)$ and the structure factor $S(k,t)$. Both quantities were spherically averaged as the system is isotropic. We found that the RFXYM in $d=2,3$ obeys dynamical scaling of $C(r,t)$ and $S(k,t)$. Further, the scaling functions are independent of the disorder amplitude, demonstrating their universality.

We also investigated the autocorrelation function $A(t,t_w)$ to study aging and the superuniversality (SU) property. In $d=2$, $A(t,t_w)$ shows a {\it non-simple} scaling with an aging exponent $B > 0$, whereas a {\it simple} scaling is observed for $d=3$. The autocorrelation functions in both $d=2,3$ exhibit a neat data collapse in the aging regime where $L(t)/L(t_w) \gg 1$. We probed this regime in the large-$t$ limit where the scaling function is governed by a power-law decay: $A(t,t_w) \sim [L(t)/L(t_w)]^{-\lambda_C}$, with $\lambda_C$ being the autocorrelation exponent. In $d=2$, the values of $\lambda_C$ for different $\Delta$ obey the bounds $d/2 \leq \lambda_C < d$ \cite{fisher1988nonequilibrium,huse1989remanent,yeung1996bounds}, while in $d=3$ the lower bound is found to be violated. Such a violation has also been reported earlier for the dilute Ising model \cite{paul2007superaging}. However, we do not have any explanation as to why only the $d=3$ RFXYM showed this violation. Most importantly, we found that the scaling functions for aging in $d=2,3$ are strongly dependent on $\Delta$, demonstrating that the SU property is not obeyed.

Before ending, it is relevant to ask about future directions in this area. The numerical work of our group, and that of other groups, has clearly demonstrated that domain growth in disordered systems generically has two regimes. \\
(a) At early times, the growth obeys a power law with a disorder-dependent exponent. This regime is well described by barriers which depend logarithmically on the domain size, $E_B(L) \sim \ln (1+aL)$, where $a$ is a constant \cite{Paul_2004,ppr05}. \\
(b) At later times, the growth shows a logarithmic behavior $L \sim (\ln t)^{1/\psi}$. Huse and Henley \cite{hh85} have argued that this results from energy barriers which scale as a power of the domain size, $E_B(L) \sim L^\psi$. \\
To the best of our knowledge, there are no reliable arguments for either of these barrier-dependences. In our opinion, the outstanding problem in this area is the formulation of a microscopic theory which derives such barriers from models like the RFIM, RBIM, RFXYM, RBXYM, etc. Our numerical understanding of these problems is now quite comprehensive. However, some analytical insights would be very welcome. It is our hope that the numerical results presented here would inspire further interest in these problems.

\subsection*{Acknowledgments}

MK would like to acknowledge the support of the Royal Society--SERB Newton International fellowship NIF/R1/180386.

\newpage
\appendix

\section{Vector Plots for Typical $d=2$ RFXYM Spin Configurations}
\label{vector_plot}

\begin{figure}[h!]
\centering
\includegraphics[width=0.98\linewidth]{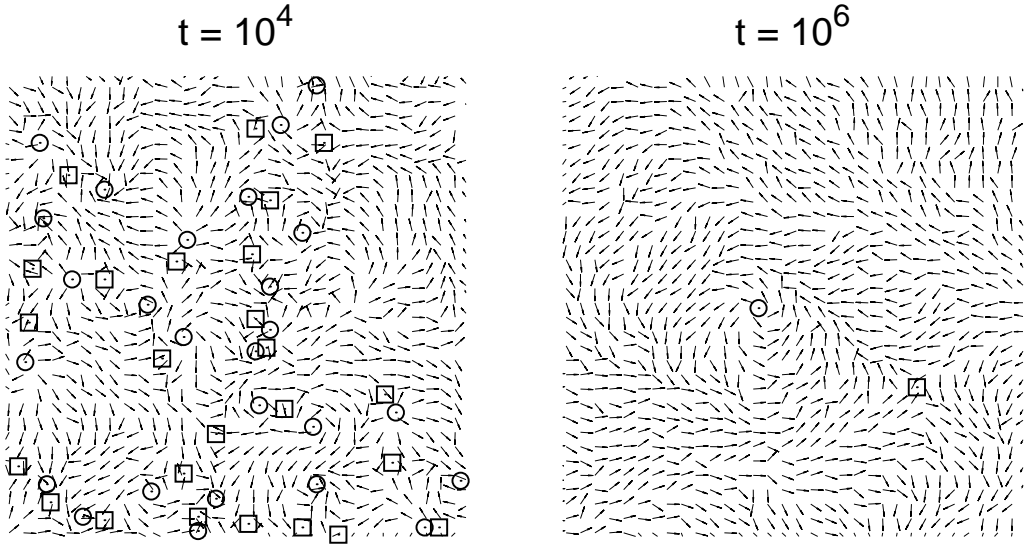}
\caption{Vector plots for the $d=2$ RFXYM with $\Delta = 0.1$ and $T=0.2$, at different times. At each lattice site $i$, a vector corresponding to spin $\mathbf{S}_i = (\cos \theta_i,\sin \theta_i)$ is drawn. For a better view, we show only a $128^2$ portion of the $1024^2$ lattice. The circles and squares denote vortices and anti-vortices, respectively.}
\label{domain}
\end{figure}
Fig.~\ref{domain} shows the typical vector plots for the $d=2$ RFXYM with $\Delta = 0.1$ at different times (in MCS) after a quench to temperature $T=0.2$. When the length scale is greater or the defect density is lower, the system shows larger regions (located at defect boundaries) where the spins orient in nearly the same direction.

\newpage

\bibliography{rfxydg}

%merlin.mbs apsrev4-1.bst 2010-07-25 4.21a (PWD, AO, DPC) hacked
%Control: key (0)
%Control: author (8) initials jnrlst
%Control: editor formatted (1) identically to author
%Control: production of article title (-1) disabled
%Control: page (0) single
%Control: year (1) truncated
%Control: production of eprint (0) enabled
\begin{thebibliography}{73}%
\makeatletter
\providecommand \@ifxundefined [1]{%
 \@ifx{#1\undefined}
}%
\providecommand \@ifnum [1]{%
 \ifnum #1\expandafter \@firstoftwo
 \else \expandafter \@secondoftwo
 \fi
}%
\providecommand \@ifx [1]{%
 \ifx #1\expandafter \@firstoftwo
 \else \expandafter \@secondoftwo
 \fi
}%
\providecommand \natexlab [1]{#1}%
\providecommand \enquote  [1]{``#1''}%
\providecommand \bibnamefont  [1]{#1}%
\providecommand \bibfnamefont [1]{#1}%
\providecommand \citenamefont [1]{#1}%
\providecommand \href@noop [0]{\@secondoftwo}%
\providecommand \href [0]{\begingroup \@sanitize@url \@href}%
\providecommand \@href[1]{\@@startlink{#1}\@@href}%
\providecommand \@@href[1]{\endgroup#1\@@endlink}%
\providecommand \@sanitize@url [0]{\catcode `\\12\catcode `\$12\catcode
  `\&12\catcode `\#12\catcode `\^12\catcode `\_12\catcode `\%12\relax}%
\providecommand \@@startlink[1]{}%
\providecommand \@@endlink[0]{}%
\providecommand \url  [0]{\begingroup\@sanitize@url \@url }%
\providecommand \@url [1]{\endgroup\@href {#1}{\urlprefix }}%
\providecommand \urlprefix  [0]{URL }%
\providecommand \Eprint [0]{\href }%
\providecommand \doibase [0]{http://dx.doi.org/}%
\providecommand \selectlanguage [0]{\@gobble}%
\providecommand \bibinfo  [0]{\@secondoftwo}%
\providecommand \bibfield  [0]{\@secondoftwo}%
\providecommand \translation [1]{[#1]}%
\providecommand \BibitemOpen [0]{}%
\providecommand \bibitemStop [0]{}%
\providecommand \bibitemNoStop [0]{.\EOS\space}%
\providecommand \EOS [0]{\spacefactor3000\relax}%
\providecommand \BibitemShut  [1]{\csname bibitem#1\endcsname}%
\let\auto@bib@innerbib\@empty
%</preamble>
\bibitem [{\citenamefont {Larkin}\ and\ \citenamefont
  {Ovchinnikov}(1979)}]{larkin1979pinning}%
  \BibitemOpen
  \bibfield  {author} {\bibinfo {author} {\bibfnamefont {A.}~\bibnamefont
  {Larkin}}\ and\ \bibinfo {author} {\bibfnamefont {Y.~N.}\ \bibnamefont
  {Ovchinnikov}},\ }\href@noop {} {\bibfield  {journal} {\bibinfo  {journal}
  {Journal of Low Temperature Physics}\ }\textbf {\bibinfo {volume} {34}},\
  \bibinfo {pages} {409} (\bibinfo {year} {1979})}\BibitemShut {NoStop}%
\bibitem [{\citenamefont {Larkin}(1970)}]{larkin1970effect}%
  \BibitemOpen
  \bibfield  {author} {\bibinfo {author} {\bibfnamefont {A.~I.}\ \bibnamefont
  {Larkin}},\ }\href@noop {} {\bibfield  {journal} {\bibinfo  {journal} {JETP}\
  }\textbf {\bibinfo {volume} {31}},\ \bibinfo {pages} {784} (\bibinfo {year}
  {1970})}\BibitemShut {NoStop}%
\bibitem [{\citenamefont {Nattermann}\ and\ \citenamefont
  {Scheidl}(2000)}]{nattermann2000vortex}%
  \BibitemOpen
  \bibfield  {author} {\bibinfo {author} {\bibfnamefont {T.}~\bibnamefont
  {Nattermann}}\ and\ \bibinfo {author} {\bibfnamefont {S.}~\bibnamefont
  {Scheidl}},\ }\href@noop {} {\bibfield  {journal} {\bibinfo  {journal} {Adv.
  Phys.}\ }\textbf {\bibinfo {volume} {49}},\ \bibinfo {pages} {607} (\bibinfo
  {year} {2000})}\BibitemShut {NoStop}%
\bibitem [{\citenamefont {Fukuyama}\ and\ \citenamefont
  {Lee}(1978)}]{fukuyama1978dynamics}%
  \BibitemOpen
  \bibfield  {author} {\bibinfo {author} {\bibfnamefont {H.}~\bibnamefont
  {Fukuyama}}\ and\ \bibinfo {author} {\bibfnamefont {P.~A.}\ \bibnamefont
  {Lee}},\ }\href@noop {} {\bibfield  {journal} {\bibinfo  {journal} {Phys.
  Rev. B}\ }\textbf {\bibinfo {volume} {17}},\ \bibinfo {pages} {535} (\bibinfo
  {year} {1978})}\BibitemShut {NoStop}%
\bibitem [{\citenamefont {Lee}\ and\ \citenamefont
  {Rice}(1979)}]{lee1979electric}%
  \BibitemOpen
  \bibfield  {author} {\bibinfo {author} {\bibfnamefont {P.}~\bibnamefont
  {Lee}}\ and\ \bibinfo {author} {\bibfnamefont {T.}~\bibnamefont {Rice}},\
  }\href@noop {} {\bibfield  {journal} {\bibinfo  {journal} {Phys. Rev. B}\
  }\textbf {\bibinfo {volume} {19}},\ \bibinfo {pages} {3970} (\bibinfo {year}
  {1979})}\BibitemShut {NoStop}%
\bibitem [{\citenamefont {Sham}\ and\ \citenamefont
  {Patton}(1976)}]{sham1976effect}%
  \BibitemOpen
  \bibfield  {author} {\bibinfo {author} {\bibfnamefont {L.}~\bibnamefont
  {Sham}}\ and\ \bibinfo {author} {\bibfnamefont {B.~R.}\ \bibnamefont
  {Patton}},\ }\href@noop {} {\bibfield  {journal} {\bibinfo  {journal} {Phys.
  Rev. B}\ }\textbf {\bibinfo {volume} {13}},\ \bibinfo {pages} {3151}
  (\bibinfo {year} {1976})}\BibitemShut {NoStop}%
\bibitem [{\citenamefont {Efetov}\ and\ \citenamefont
  {Larkin}(1977)}]{efetov1977charge}%
  \BibitemOpen
  \bibfield  {author} {\bibinfo {author} {\bibfnamefont {K.}~\bibnamefont
  {Efetov}}\ and\ \bibinfo {author} {\bibfnamefont {A.}~\bibnamefont
  {Larkin}},\ }\href@noop {} {\bibfield  {journal} {\bibinfo  {journal} {Sov.
  Phys. JETP}\ }\textbf {\bibinfo {volume} {45}},\ \bibinfo {pages} {2350}
  (\bibinfo {year} {1977})}\BibitemShut {NoStop}%
\bibitem [{\citenamefont {Iannacchione}\ \emph {et~al.}(1993)\citenamefont
  {Iannacchione}, \citenamefont {Crawford}, \citenamefont {{\v{Z}}umer},
  \citenamefont {Doane},\ and\ \citenamefont
  {Finotello}}]{iannacchione1993randomly}%
  \BibitemOpen
  \bibfield  {author} {\bibinfo {author} {\bibfnamefont {G.}~\bibnamefont
  {Iannacchione}}, \bibinfo {author} {\bibfnamefont {G.}~\bibnamefont
  {Crawford}}, \bibinfo {author} {\bibfnamefont {S.}~\bibnamefont
  {{\v{Z}}umer}}, \bibinfo {author} {\bibfnamefont {J.}~\bibnamefont {Doane}},
  \ and\ \bibinfo {author} {\bibfnamefont {D.}~\bibnamefont {Finotello}},\
  }\href@noop {} {\bibfield  {journal} {\bibinfo  {journal} {Phys. Rev. Lett.}\
  }\textbf {\bibinfo {volume} {71}},\ \bibinfo {pages} {2595} (\bibinfo {year}
  {1993})}\BibitemShut {NoStop}%
\bibitem [{\citenamefont {Sellmyer}\ and\ \citenamefont
  {O’Shea}(1992)}]{sellmyer1992random}%
  \BibitemOpen
  \bibfield  {author} {\bibinfo {author} {\bibfnamefont {D.}~\bibnamefont
  {Sellmyer}}\ and\ \bibinfo {author} {\bibfnamefont {M.}~\bibnamefont
  {O’Shea}},\ }\href@noop {} {\bibfield  {journal} {\bibinfo  {journal}
  {Recent Progress in Random Magnets, World Scientific, Singapore}\ ,\ \bibinfo
  {pages} {71}} (\bibinfo {year} {1992})}\BibitemShut {NoStop}%
\bibitem [{\citenamefont {Berezinskii}(1971)}]{berezinskii1971destruction}%
  \BibitemOpen
  \bibfield  {author} {\bibinfo {author} {\bibfnamefont {V.}~\bibnamefont
  {Berezinskii}},\ }\href@noop {} {\bibfield  {journal} {\bibinfo  {journal}
  {Sov. Phys. JETP}\ }\textbf {\bibinfo {volume} {32}},\ \bibinfo {pages} {493}
  (\bibinfo {year} {1971})}\BibitemShut {NoStop}%
\bibitem [{\citenamefont {Kosterlitz}\ and\ \citenamefont
  {Thouless}(1973)}]{kosterlitz1973ordering}%
  \BibitemOpen
  \bibfield  {author} {\bibinfo {author} {\bibfnamefont {J.~M.}\ \bibnamefont
  {Kosterlitz}}\ and\ \bibinfo {author} {\bibfnamefont {D.~J.}\ \bibnamefont
  {Thouless}},\ }\href@noop {} {\bibfield  {journal} {\bibinfo  {journal} {J.
  Phys. C}\ }\textbf {\bibinfo {volume} {6}},\ \bibinfo {pages} {1181}
  (\bibinfo {year} {1973})}\BibitemShut {NoStop}%
\bibitem [{\citenamefont {Kosterlitz}(1974)}]{kosterlitz1974critical}%
  \BibitemOpen
  \bibfield  {author} {\bibinfo {author} {\bibfnamefont {J.}~\bibnamefont
  {Kosterlitz}},\ }\href@noop {} {\bibfield  {journal} {\bibinfo  {journal} {J.
  Phys. C}\ }\textbf {\bibinfo {volume} {7}},\ \bibinfo {pages} {1046}
  (\bibinfo {year} {1974})}\BibitemShut {NoStop}%
\bibitem [{\citenamefont {Imry}\ and\ \citenamefont
  {Ma}(1975)}]{imry1975random}%
  \BibitemOpen
  \bibfield  {author} {\bibinfo {author} {\bibfnamefont {Y.}~\bibnamefont
  {Imry}}\ and\ \bibinfo {author} {\bibfnamefont {S.-k.}\ \bibnamefont {Ma}},\
  }\href@noop {} {\bibfield  {journal} {\bibinfo  {journal} {Phys. Rev. Lett.}\
  }\textbf {\bibinfo {volume} {35}},\ \bibinfo {pages} {1399} (\bibinfo {year}
  {1975})}\BibitemShut {NoStop}%
\bibitem [{\citenamefont {Aizenman}\ and\ \citenamefont
  {Wehr}(1989)}]{aizenman1989rounding}%
  \BibitemOpen
  \bibfield  {author} {\bibinfo {author} {\bibfnamefont {M.}~\bibnamefont
  {Aizenman}}\ and\ \bibinfo {author} {\bibfnamefont {J.}~\bibnamefont
  {Wehr}},\ }\href@noop {} {\bibfield  {journal} {\bibinfo  {journal} {Phys.
  Rev. Lett.}\ }\textbf {\bibinfo {volume} {62}},\ \bibinfo {pages} {2503}
  (\bibinfo {year} {1989})}\BibitemShut {NoStop}%
\bibitem [{\citenamefont {Aizenman}\ and\ \citenamefont
  {Wehr}(1990)}]{aizenman1990rounding}%
  \BibitemOpen
  \bibfield  {author} {\bibinfo {author} {\bibfnamefont {M.}~\bibnamefont
  {Aizenman}}\ and\ \bibinfo {author} {\bibfnamefont {J.}~\bibnamefont
  {Wehr}},\ }\href@noop {} {\bibfield  {journal} {\bibinfo  {journal} {Commun.
  Math. Phys.}\ }\textbf {\bibinfo {volume} {130}},\ \bibinfo {pages} {489}
  (\bibinfo {year} {1990})}\BibitemShut {NoStop}%
\bibitem [{\citenamefont {Korshunov}(1993)}]{korshunov1993replica}%
  \BibitemOpen
  \bibfield  {author} {\bibinfo {author} {\bibfnamefont {S.}~\bibnamefont
  {Korshunov}},\ }\href@noop {} {\bibfield  {journal} {\bibinfo  {journal}
  {Phys. Rev. B}\ }\textbf {\bibinfo {volume} {48}},\ \bibinfo {pages} {3969}
  (\bibinfo {year} {1993})}\BibitemShut {NoStop}%
\bibitem [{\citenamefont {Garel}\ \emph {et~al.}(1996)\citenamefont {Garel},
  \citenamefont {Iori},\ and\ \citenamefont {Orland}}]{garel1996variational}%
  \BibitemOpen
  \bibfield  {author} {\bibinfo {author} {\bibfnamefont {T.}~\bibnamefont
  {Garel}}, \bibinfo {author} {\bibfnamefont {G.}~\bibnamefont {Iori}}, \ and\
  \bibinfo {author} {\bibfnamefont {H.}~\bibnamefont {Orland}},\ }\href@noop {}
  {\bibfield  {journal} {\bibinfo  {journal} {Phys. Rev. B}\ }\textbf {\bibinfo
  {volume} {53}},\ \bibinfo {pages} {R2941} (\bibinfo {year}
  {1996})}\BibitemShut {NoStop}%
\bibitem [{\citenamefont {Feldman}(2001)}]{feldman2001quasi}%
  \BibitemOpen
  \bibfield  {author} {\bibinfo {author} {\bibfnamefont {D.}~\bibnamefont
  {Feldman}},\ }\href@noop {} {\bibfield  {journal} {\bibinfo  {journal} {Int.
  J. Mod. Phys. B}\ }\textbf {\bibinfo {volume} {15}},\ \bibinfo {pages} {2945}
  (\bibinfo {year} {2001})}\BibitemShut {NoStop}%
\bibitem [{\citenamefont {Giamarchi}\ and\ \citenamefont
  {Le~Doussal}(1995)}]{giamarchi1995elastic}%
  \BibitemOpen
  \bibfield  {author} {\bibinfo {author} {\bibfnamefont {T.}~\bibnamefont
  {Giamarchi}}\ and\ \bibinfo {author} {\bibfnamefont {P.}~\bibnamefont
  {Le~Doussal}},\ }\href@noop {} {\bibfield  {journal} {\bibinfo  {journal}
  {Phys. Rev. B}\ }\textbf {\bibinfo {volume} {52}},\ \bibinfo {pages} {1242}
  (\bibinfo {year} {1995})}\BibitemShut {NoStop}%
\bibitem [{\citenamefont {Fisher}(1997)}]{fisher1997stability}%
  \BibitemOpen
  \bibfield  {author} {\bibinfo {author} {\bibfnamefont {D.~S.}\ \bibnamefont
  {Fisher}},\ }\href@noop {} {\bibfield  {journal} {\bibinfo  {journal} {Phys.
  Rev. Lett.}\ }\textbf {\bibinfo {volume} {78}},\ \bibinfo {pages} {1964}
  (\bibinfo {year} {1997})}\BibitemShut {NoStop}%
\bibitem [{\citenamefont {Gingras}\ and\ \citenamefont
  {Huse}(1996)}]{gingras1996topological}%
  \BibitemOpen
  \bibfield  {author} {\bibinfo {author} {\bibfnamefont {M.~J.}\ \bibnamefont
  {Gingras}}\ and\ \bibinfo {author} {\bibfnamefont {D.~A.}\ \bibnamefont
  {Huse}},\ }\href@noop {} {\bibfield  {journal} {\bibinfo  {journal} {Phys.
  Rev. B}\ }\textbf {\bibinfo {volume} {53}},\ \bibinfo {pages} {15193}
  (\bibinfo {year} {1996})}\BibitemShut {NoStop}%
\bibitem [{\citenamefont {Fisch}(1997)}]{fisch1997power}%
  \BibitemOpen
  \bibfield  {author} {\bibinfo {author} {\bibfnamefont {R.}~\bibnamefont
  {Fisch}},\ }\href@noop {} {\bibfield  {journal} {\bibinfo  {journal} {Phys.
  Rev. B}\ }\textbf {\bibinfo {volume} {55}},\ \bibinfo {pages} {8211}
  (\bibinfo {year} {1997})}\BibitemShut {NoStop}%
\bibitem [{\citenamefont {Fisch}(2000)}]{fisch2000random}%
  \BibitemOpen
  \bibfield  {author} {\bibinfo {author} {\bibfnamefont {R.}~\bibnamefont
  {Fisch}},\ }\href@noop {} {\bibfield  {journal} {\bibinfo  {journal} {Phys.
  Rev. B}\ }\textbf {\bibinfo {volume} {62}},\ \bibinfo {pages} {361} (\bibinfo
  {year} {2000})}\BibitemShut {NoStop}%
\bibitem [{\citenamefont {Fisch}(2007)}]{fisch2007structure}%
  \BibitemOpen
  \bibfield  {author} {\bibinfo {author} {\bibfnamefont {R.}~\bibnamefont
  {Fisch}},\ }\href@noop {} {\bibfield  {journal} {\bibinfo  {journal} {Phys.
  Rev. B}\ }\textbf {\bibinfo {volume} {76}},\ \bibinfo {pages} {214435}
  (\bibinfo {year} {2007})}\BibitemShut {NoStop}%
\bibitem [{\citenamefont {Tissier}\ and\ \citenamefont
  {Tarjus}(2006{\natexlab{a}})}]{tissier2006unified}%
  \BibitemOpen
  \bibfield  {author} {\bibinfo {author} {\bibfnamefont {M.}~\bibnamefont
  {Tissier}}\ and\ \bibinfo {author} {\bibfnamefont {G.}~\bibnamefont
  {Tarjus}},\ }\href@noop {} {\bibfield  {journal} {\bibinfo  {journal} {Phys.
  Rev. Lett.}\ }\textbf {\bibinfo {volume} {96}},\ \bibinfo {pages} {087202}
  (\bibinfo {year} {2006}{\natexlab{a}})}\BibitemShut {NoStop}%
\bibitem [{\citenamefont {Tissier}\ and\ \citenamefont
  {Tarjus}(2006{\natexlab{b}})}]{tissier2006two}%
  \BibitemOpen
  \bibfield  {author} {\bibinfo {author} {\bibfnamefont {M.}~\bibnamefont
  {Tissier}}\ and\ \bibinfo {author} {\bibfnamefont {G.}~\bibnamefont
  {Tarjus}},\ }\href@noop {} {\bibfield  {journal} {\bibinfo  {journal} {Phys.
  Rev. B}\ }\textbf {\bibinfo {volume} {74}},\ \bibinfo {pages} {214419}
  (\bibinfo {year} {2006}{\natexlab{b}})}\BibitemShut {NoStop}%
\bibitem [{\citenamefont {Tarjus}\ and\ \citenamefont
  {Tissier}(2020)}]{tarjus2020random}%
  \BibitemOpen
  \bibfield  {author} {\bibinfo {author} {\bibfnamefont {G.}~\bibnamefont
  {Tarjus}}\ and\ \bibinfo {author} {\bibfnamefont {M.}~\bibnamefont
  {Tissier}},\ }\href@noop {} {\bibfield  {journal} {\bibinfo  {journal} {Eur.
  Phys. J. B}\ }\textbf {\bibinfo {volume} {93}},\ \bibinfo {pages} {1}
  (\bibinfo {year} {2020})}\BibitemShut {NoStop}%
\bibitem [{\citenamefont {Garanin}\ \emph {et~al.}(2013)\citenamefont
  {Garanin}, \citenamefont {Chudnovsky},\ and\ \citenamefont
  {Proctor}}]{garanin2013random}%
  \BibitemOpen
  \bibfield  {author} {\bibinfo {author} {\bibfnamefont {D.}~\bibnamefont
  {Garanin}}, \bibinfo {author} {\bibfnamefont {E.}~\bibnamefont {Chudnovsky}},
  \ and\ \bibinfo {author} {\bibfnamefont {T.}~\bibnamefont {Proctor}},\
  }\href@noop {} {\bibfield  {journal} {\bibinfo  {journal} {Phys. Rev. B}\
  }\textbf {\bibinfo {volume} {88}},\ \bibinfo {pages} {224418} (\bibinfo
  {year} {2013})}\BibitemShut {NoStop}%
\bibitem [{\citenamefont {Ray}\ and\ \citenamefont
  {Moore}(1992)}]{ray1992chirality}%
  \BibitemOpen
  \bibfield  {author} {\bibinfo {author} {\bibfnamefont {P.}~\bibnamefont
  {Ray}}\ and\ \bibinfo {author} {\bibfnamefont {M.}~\bibnamefont {Moore}},\
  }\href@noop {} {\bibfield  {journal} {\bibinfo  {journal} {Phys. Rev. B}\
  }\textbf {\bibinfo {volume} {45}},\ \bibinfo {pages} {5361} (\bibinfo {year}
  {1992})}\BibitemShut {NoStop}%
\bibitem [{\citenamefont {Tang}\ \emph {et~al.}(2015)\citenamefont {Tang},
  \citenamefont {Iyer},\ and\ \citenamefont {Rigol}}]{tang2015thermodynamics}%
  \BibitemOpen
  \bibfield  {author} {\bibinfo {author} {\bibfnamefont {B.}~\bibnamefont
  {Tang}}, \bibinfo {author} {\bibfnamefont {D.}~\bibnamefont {Iyer}}, \ and\
  \bibinfo {author} {\bibfnamefont {M.}~\bibnamefont {Rigol}},\ }\href@noop {}
  {\bibfield  {journal} {\bibinfo  {journal} {Phys. Rev. B}\ }\textbf {\bibinfo
  {volume} {91}},\ \bibinfo {pages} {174413} (\bibinfo {year}
  {2015})}\BibitemShut {NoStop}%
\bibitem [{\citenamefont {Kumar}\ \emph {et~al.}(2017)\citenamefont {Kumar},
  \citenamefont {Chatterjee}, \citenamefont {Paul},\ and\ \citenamefont
  {Puri}}]{kumar2017ordering}%
  \BibitemOpen
  \bibfield  {author} {\bibinfo {author} {\bibfnamefont {M.}~\bibnamefont
  {Kumar}}, \bibinfo {author} {\bibfnamefont {S.}~\bibnamefont {Chatterjee}},
  \bibinfo {author} {\bibfnamefont {R.}~\bibnamefont {Paul}}, \ and\ \bibinfo
  {author} {\bibfnamefont {S.}~\bibnamefont {Puri}},\ }\href@noop {} {\bibfield
   {journal} {\bibinfo  {journal} {Phys. Rev. E}\ }\textbf {\bibinfo {volume}
  {96}},\ \bibinfo {pages} {042127} (\bibinfo {year} {2017})}\BibitemShut
  {NoStop}%
\bibitem [{\citenamefont {Li}\ \emph {et~al.}(1996)\citenamefont {Li},
  \citenamefont {Nattermann}, \citenamefont {Rieger},\ and\ \citenamefont
  {Schwartz}}]{li1996vortex}%
  \BibitemOpen
  \bibfield  {author} {\bibinfo {author} {\bibfnamefont {M.~S.}\ \bibnamefont
  {Li}}, \bibinfo {author} {\bibfnamefont {T.}~\bibnamefont {Nattermann}},
  \bibinfo {author} {\bibfnamefont {H.}~\bibnamefont {Rieger}}, \ and\ \bibinfo
  {author} {\bibfnamefont {M.}~\bibnamefont {Schwartz}},\ }\href@noop {}
  {\bibfield  {journal} {\bibinfo  {journal} {Phys. Rev. B}\ }\textbf {\bibinfo
  {volume} {54}},\ \bibinfo {pages} {16024} (\bibinfo {year}
  {1996})}\BibitemShut {NoStop}%
\bibitem [{\citenamefont {Leonel}\ \emph {et~al.}(2003)\citenamefont {Leonel},
  \citenamefont {Coura}, \citenamefont {Pereira}, \citenamefont {M{\'o}l},\
  and\ \citenamefont {Costa}}]{leonel2003monte}%
  \BibitemOpen
  \bibfield  {author} {\bibinfo {author} {\bibfnamefont {S.}~\bibnamefont
  {Leonel}}, \bibinfo {author} {\bibfnamefont {P.~Z.}\ \bibnamefont {Coura}},
  \bibinfo {author} {\bibfnamefont {A.}~\bibnamefont {Pereira}}, \bibinfo
  {author} {\bibfnamefont {L.}~\bibnamefont {M{\'o}l}}, \ and\ \bibinfo
  {author} {\bibfnamefont {B.}~\bibnamefont {Costa}},\ }\href@noop {}
  {\bibfield  {journal} {\bibinfo  {journal} {Phys. Rev. B}\ }\textbf {\bibinfo
  {volume} {67}},\ \bibinfo {pages} {104426} (\bibinfo {year}
  {2003})}\BibitemShut {NoStop}%
\bibitem [{\citenamefont {Wysin}\ \emph {et~al.}(2005)\citenamefont {Wysin},
  \citenamefont {Pereira}, \citenamefont {Marques}, \citenamefont {Leonel},\
  and\ \citenamefont {Coura}}]{wysin2005extinction}%
  \BibitemOpen
  \bibfield  {author} {\bibinfo {author} {\bibfnamefont {G.}~\bibnamefont
  {Wysin}}, \bibinfo {author} {\bibfnamefont {A.}~\bibnamefont {Pereira}},
  \bibinfo {author} {\bibfnamefont {I.}~\bibnamefont {Marques}}, \bibinfo
  {author} {\bibfnamefont {S.}~\bibnamefont {Leonel}}, \ and\ \bibinfo {author}
  {\bibfnamefont {P.}~\bibnamefont {Coura}},\ }\href@noop {} {\bibfield
  {journal} {\bibinfo  {journal} {Phys. Rev. B}\ }\textbf {\bibinfo {volume}
  {72}},\ \bibinfo {pages} {094418} (\bibinfo {year} {2005})}\BibitemShut
  {NoStop}%
\bibitem [{\citenamefont {Surungan}\ and\ \citenamefont
  {Okabe}(2005)}]{surungan2005kosterlitz}%
  \BibitemOpen
  \bibfield  {author} {\bibinfo {author} {\bibfnamefont {T.}~\bibnamefont
  {Surungan}}\ and\ \bibinfo {author} {\bibfnamefont {Y.}~\bibnamefont
  {Okabe}},\ }\href@noop {} {\bibfield  {journal} {\bibinfo  {journal} {Phys.
  Rev. B}\ }\textbf {\bibinfo {volume} {71}},\ \bibinfo {pages} {184438}
  (\bibinfo {year} {2005})}\BibitemShut {NoStop}%
\bibitem [{\citenamefont {Tagantsev}\ \emph {et~al.}(2010)\citenamefont
  {Tagantsev}, \citenamefont {Cross},\ and\ \citenamefont {Fousek}}]{tcf10}%
  \BibitemOpen
  \bibfield  {author} {\bibinfo {author} {\bibfnamefont {A.~K.}\ \bibnamefont
  {Tagantsev}}, \bibinfo {author} {\bibfnamefont {L.~E.}\ \bibnamefont
  {Cross}}, \ and\ \bibinfo {author} {\bibfnamefont {J.}~\bibnamefont
  {Fousek}},\ }\href@noop {} {\emph {\bibinfo {title} {Domains in ferroic
  crystals and thin films}}}\ (\bibinfo  {publisher} {Springer},\ \bibinfo
  {year} {2010})\BibitemShut {NoStop}%
\bibitem [{\citenamefont {Puri}(2004)}]{sp04}%
  \BibitemOpen
  \bibfield  {author} {\bibinfo {author} {\bibfnamefont {S.}~\bibnamefont
  {Puri}},\ }\href@noop {} {\bibfield  {journal} {\bibinfo  {journal} {Phase
  Transitions}\ }\textbf {\bibinfo {volume} {77}},\ \bibinfo {pages} {469}
  (\bibinfo {year} {2004})}\BibitemShut {NoStop}%
\bibitem [{\citenamefont {Yurke}\ \emph {et~al.}(1993)\citenamefont {Yurke},
  \citenamefont {Pargellis}, \citenamefont {Kovacs},\ and\ \citenamefont
  {Huse}}]{yurke1993coarsening}%
  \BibitemOpen
  \bibfield  {author} {\bibinfo {author} {\bibfnamefont {B.}~\bibnamefont
  {Yurke}}, \bibinfo {author} {\bibfnamefont {A.}~\bibnamefont {Pargellis}},
  \bibinfo {author} {\bibfnamefont {T.}~\bibnamefont {Kovacs}}, \ and\ \bibinfo
  {author} {\bibfnamefont {D.}~\bibnamefont {Huse}},\ }\href@noop {} {\bibfield
   {journal} {\bibinfo  {journal} {Phys. Rev. E}\ }\textbf {\bibinfo {volume}
  {47}},\ \bibinfo {pages} {1525} (\bibinfo {year} {1993})}\BibitemShut
  {NoStop}%
\bibitem [{\citenamefont {Kohring}\ \emph {et~al.}(1986)\citenamefont
  {Kohring}, \citenamefont {Shrock},\ and\ \citenamefont
  {Wills}}]{kohring1986role}%
  \BibitemOpen
  \bibfield  {author} {\bibinfo {author} {\bibfnamefont {G.}~\bibnamefont
  {Kohring}}, \bibinfo {author} {\bibfnamefont {R.~E.}\ \bibnamefont {Shrock}},
  \ and\ \bibinfo {author} {\bibfnamefont {P.}~\bibnamefont {Wills}},\
  }\href@noop {} {\bibfield  {journal} {\bibinfo  {journal} {Phys. Rev. Lett.}\
  }\textbf {\bibinfo {volume} {57}},\ \bibinfo {pages} {1358} (\bibinfo {year}
  {1986})}\BibitemShut {NoStop}%
\bibitem [{\citenamefont {Gottlob}\ and\ \citenamefont
  {Hasenbusch}(1993)}]{gottlob1993critical}%
  \BibitemOpen
  \bibfield  {author} {\bibinfo {author} {\bibfnamefont {A.~P.}\ \bibnamefont
  {Gottlob}}\ and\ \bibinfo {author} {\bibfnamefont {M.}~\bibnamefont
  {Hasenbusch}},\ }\href@noop {} {\bibfield  {journal} {\bibinfo  {journal}
  {Physica A}\ }\textbf {\bibinfo {volume} {201}},\ \bibinfo {pages} {593}
  (\bibinfo {year} {1993})}\BibitemShut {NoStop}%
\bibitem [{\citenamefont {Tobochnik}\ and\ \citenamefont
  {Chester}(1979)}]{tobochnik1979monte}%
  \BibitemOpen
  \bibfield  {author} {\bibinfo {author} {\bibfnamefont {J.}~\bibnamefont
  {Tobochnik}}\ and\ \bibinfo {author} {\bibfnamefont {G.}~\bibnamefont
  {Chester}},\ }\href@noop {} {\bibfield  {journal} {\bibinfo  {journal} {Phys.
  Rev. B}\ }\textbf {\bibinfo {volume} {20}},\ \bibinfo {pages} {3761}
  (\bibinfo {year} {1979})}\BibitemShut {NoStop}%
\bibitem [{\citenamefont {Fern{\'a}ndez}\ \emph {et~al.}(1986)\citenamefont
  {Fern{\'a}ndez}, \citenamefont {Ferreira},\ and\ \citenamefont
  {Stankiewicz}}]{fernandez1986critical}%
  \BibitemOpen
  \bibfield  {author} {\bibinfo {author} {\bibfnamefont {J.~F.}\ \bibnamefont
  {Fern{\'a}ndez}}, \bibinfo {author} {\bibfnamefont {M.~F.}\ \bibnamefont
  {Ferreira}}, \ and\ \bibinfo {author} {\bibfnamefont {J.}~\bibnamefont
  {Stankiewicz}},\ }\href@noop {} {\bibfield  {journal} {\bibinfo  {journal}
  {Phys. Rev. B}\ }\textbf {\bibinfo {volume} {34}},\ \bibinfo {pages} {292}
  (\bibinfo {year} {1986})}\BibitemShut {NoStop}%
\bibitem [{\citenamefont {Hasenbusch}\ and\ \citenamefont
  {Meyer}(1990)}]{hasenbusch1990critical}%
  \BibitemOpen
  \bibfield  {author} {\bibinfo {author} {\bibfnamefont {M.}~\bibnamefont
  {Hasenbusch}}\ and\ \bibinfo {author} {\bibfnamefont {S.}~\bibnamefont
  {Meyer}},\ }\href@noop {} {\bibfield  {journal} {\bibinfo  {journal} {Phys.
  Lett. B}\ }\textbf {\bibinfo {volume} {241}},\ \bibinfo {pages} {238}
  (\bibinfo {year} {1990})}\BibitemShut {NoStop}%
\bibitem [{\citenamefont {Newman}\ and\ \citenamefont
  {Barkema}(1999)}]{newman1999monte}%
  \BibitemOpen
  \bibfield  {author} {\bibinfo {author} {\bibfnamefont {M.}~\bibnamefont
  {Newman}}\ and\ \bibinfo {author} {\bibfnamefont {G.}~\bibnamefont
  {Barkema}},\ }\href@noop {} {\emph {\bibinfo {title} {Monte Carlo Methods in
  Statistical Physics}}},\ Vol.~\bibinfo {volume} {24}\ (\bibinfo  {publisher}
  {Oxford University Press: New York, USA},\ \bibinfo {year}
  {1999})\BibitemShut {NoStop}%
\bibitem [{\citenamefont {Puri}\ and\ \citenamefont
  {Wadhawan}(2009)}]{puri2009kinetics}%
  \BibitemOpen
  \bibfield  {author} {\bibinfo {author} {\bibfnamefont {S.}~\bibnamefont
  {Puri}}\ and\ \bibinfo {author} {\bibfnamefont {V.}~\bibnamefont
  {Wadhawan}},\ }\href@noop {} {\emph {\bibinfo {title} {Kinetics of Phase
  Transitions}}}\ (\bibinfo  {publisher} {CRC press},\ \bibinfo {year}
  {2009})\BibitemShut {NoStop}%
\bibitem [{\citenamefont {Bray}(1994)}]{Bray1994}%
  \BibitemOpen
  \bibfield  {author} {\bibinfo {author} {\bibfnamefont {A.}~\bibnamefont
  {Bray}},\ }\href {\doibase 10.1080/00018739400101505} {\bibfield  {journal}
  {\bibinfo  {journal} {Advances in Physics}\ }\textbf {\bibinfo {volume}
  {43}},\ \bibinfo {pages} {357} (\bibinfo {year} {1994})}\BibitemShut
  {NoStop}%
\bibitem [{\citenamefont {Hertel}\ and\ \citenamefont
  {Schneider}(2006)}]{PhysRevLett.97.177202}%
  \BibitemOpen
  \bibfield  {author} {\bibinfo {author} {\bibfnamefont {R.}~\bibnamefont
  {Hertel}}\ and\ \bibinfo {author} {\bibfnamefont {C.~M.}\ \bibnamefont
  {Schneider}},\ }\href {\doibase 10.1103/PhysRevLett.97.177202} {\bibfield
  {journal} {\bibinfo  {journal} {Phys. Rev. Lett.}\ }\textbf {\bibinfo
  {volume} {97}},\ \bibinfo {pages} {177202} (\bibinfo {year}
  {2006})}\BibitemShut {NoStop}%
\bibitem [{\citenamefont {Bray}\ and\ \citenamefont
  {Puri}(1991)}]{bray1991asymptotic}%
  \BibitemOpen
  \bibfield  {author} {\bibinfo {author} {\bibfnamefont {A.}~\bibnamefont
  {Bray}}\ and\ \bibinfo {author} {\bibfnamefont {S.}~\bibnamefont {Puri}},\
  }\href@noop {} {\bibfield  {journal} {\bibinfo  {journal} {Phys. Rev. Lett.}\
  }\textbf {\bibinfo {volume} {67}},\ \bibinfo {pages} {2670} (\bibinfo {year}
  {1991})}\BibitemShut {NoStop}%
\bibitem [{\citenamefont {Toyoki}(1992)}]{toyoki1992structure}%
  \BibitemOpen
  \bibfield  {author} {\bibinfo {author} {\bibfnamefont {H.}~\bibnamefont
  {Toyoki}},\ }\href@noop {} {\bibfield  {journal} {\bibinfo  {journal} {Phys.
  Rev. B}\ }\textbf {\bibinfo {volume} {45}},\ \bibinfo {pages} {1965}
  (\bibinfo {year} {1992})}\BibitemShut {NoStop}%
\bibitem [{\citenamefont {Porod}(1982)}]{porod1982small}%
  \BibitemOpen
  \bibfield  {author} {\bibinfo {author} {\bibfnamefont {G.}~\bibnamefont
  {Porod}},\ }\href@noop {} {\bibfield  {journal} {\bibinfo  {journal} {in O.
  Glatter and O. Kratky (eds.), Academic Press, London}\ } (\bibinfo {year}
  {1982})}\BibitemShut {NoStop}%
\bibitem [{\citenamefont {Oono}\ and\ \citenamefont
  {Puri}(1988)}]{oono1988large}%
  \BibitemOpen
  \bibfield  {author} {\bibinfo {author} {\bibfnamefont {Y.}~\bibnamefont
  {Oono}}\ and\ \bibinfo {author} {\bibfnamefont {S.}~\bibnamefont {Puri}},\
  }\href@noop {} {\bibfield  {journal} {\bibinfo  {journal} {Mod. Phys. Lett.
  B}\ }\textbf {\bibinfo {volume} {2}},\ \bibinfo {pages} {861} (\bibinfo
  {year} {1988})}\BibitemShut {NoStop}%
\bibitem [{\citenamefont {Zannetti}(2014)}]{zannetti2014aging}%
  \BibitemOpen
  \bibfield  {author} {\bibinfo {author} {\bibfnamefont {M.}~\bibnamefont
  {Zannetti}},\ }\href@noop {} {\bibfield  {journal} {\bibinfo  {journal}
  {arXiv preprint arXiv:1412.4670}\ } (\bibinfo {year} {2014})}\BibitemShut
  {NoStop}%
\bibitem [{\citenamefont {Kumar}\ \emph {et~al.}(2020)\citenamefont {Kumar},
  \citenamefont {Corberi}, \citenamefont {Lippiello},\ and\ \citenamefont
  {Puri}}]{kumar2020growth}%
  \BibitemOpen
  \bibfield  {author} {\bibinfo {author} {\bibfnamefont {M.}~\bibnamefont
  {Kumar}}, \bibinfo {author} {\bibfnamefont {F.}~\bibnamefont {Corberi}},
  \bibinfo {author} {\bibfnamefont {E.}~\bibnamefont {Lippiello}}, \ and\
  \bibinfo {author} {\bibfnamefont {S.}~\bibnamefont {Puri}},\ }\href@noop {}
  {\bibfield  {journal} {\bibinfo  {journal} {Eur. Phys. J. B}\ }\textbf
  {\bibinfo {volume} {93}},\ \bibinfo {pages} {1} (\bibinfo {year}
  {2020})}\BibitemShut {NoStop}%
\bibitem [{\citenamefont {Henkel}\ and\ \citenamefont
  {Pleimling}(2011)}]{henkel2011non}%
  \BibitemOpen
  \bibfield  {author} {\bibinfo {author} {\bibfnamefont {M.}~\bibnamefont
  {Henkel}}\ and\ \bibinfo {author} {\bibfnamefont {M.}~\bibnamefont
  {Pleimling}},\ }\href {https://books.google.co.uk/books?id=AiofeEteLVcC}
  {\emph {\bibinfo {title} {Non-Equilibrium Phase Transitions: Volume 2: Ageing
  and Dynamical Scaling Far from Equilibrium}}},\ Theoretical and Mathematical
  Physics\ (\bibinfo  {publisher} {Springer Netherlands},\ \bibinfo {year}
  {2011})\BibitemShut {NoStop}%
\bibitem [{\citenamefont {Fisher}\ and\ \citenamefont
  {Huse}(1988)}]{fisher1988nonequilibrium}%
  \BibitemOpen
  \bibfield  {author} {\bibinfo {author} {\bibfnamefont {D.~S.}\ \bibnamefont
  {Fisher}}\ and\ \bibinfo {author} {\bibfnamefont {D.~A.}\ \bibnamefont
  {Huse}},\ }\href@noop {} {\bibfield  {journal} {\bibinfo  {journal} {Phys.
  Rev. B}\ }\textbf {\bibinfo {volume} {38}},\ \bibinfo {pages} {373} (\bibinfo
  {year} {1988})}\BibitemShut {NoStop}%
\bibitem [{\citenamefont {Lippiello}\ \emph {et~al.}(2010)\citenamefont
  {Lippiello}, \citenamefont {Mukherjee}, \citenamefont {Puri},\ and\
  \citenamefont {Zannetti}}]{lippiello2010scaling}%
  \BibitemOpen
  \bibfield  {author} {\bibinfo {author} {\bibfnamefont {E.}~\bibnamefont
  {Lippiello}}, \bibinfo {author} {\bibfnamefont {A.}~\bibnamefont
  {Mukherjee}}, \bibinfo {author} {\bibfnamefont {S.}~\bibnamefont {Puri}}, \
  and\ \bibinfo {author} {\bibfnamefont {M.}~\bibnamefont {Zannetti}},\
  }\href@noop {} {\bibfield  {journal} {\bibinfo  {journal} {Europhys. Lett.}\
  }\textbf {\bibinfo {volume} {90}},\ \bibinfo {pages} {46006} (\bibinfo {year}
  {2010})}\BibitemShut {NoStop}%
\bibitem [{\citenamefont {Corberi}\ \emph {et~al.}(2011)\citenamefont
  {Corberi}, \citenamefont {Lippiello}, \citenamefont {Mukherjee},
  \citenamefont {Puri},\ and\ \citenamefont {Zannetti}}]{corberi2011growth}%
  \BibitemOpen
  \bibfield  {author} {\bibinfo {author} {\bibfnamefont {F.}~\bibnamefont
  {Corberi}}, \bibinfo {author} {\bibfnamefont {E.}~\bibnamefont {Lippiello}},
  \bibinfo {author} {\bibfnamefont {A.}~\bibnamefont {Mukherjee}}, \bibinfo
  {author} {\bibfnamefont {S.}~\bibnamefont {Puri}}, \ and\ \bibinfo {author}
  {\bibfnamefont {M.}~\bibnamefont {Zannetti}},\ }\href@noop {} {\bibfield
  {journal} {\bibinfo  {journal} {J Stat. Mech.: Theory Exp.}\ }\textbf
  {\bibinfo {volume} {2011}},\ \bibinfo {pages} {P03016} (\bibinfo {year}
  {2011})}\BibitemShut {NoStop}%
\bibitem [{\citenamefont {Corberi}\ \emph {et~al.}(2012)\citenamefont
  {Corberi}, \citenamefont {Lippiello}, \citenamefont {Mukherjee},
  \citenamefont {Puri},\ and\ \citenamefont {Zannetti}}]{corberi2012crossover}%
  \BibitemOpen
  \bibfield  {author} {\bibinfo {author} {\bibfnamefont {F.}~\bibnamefont
  {Corberi}}, \bibinfo {author} {\bibfnamefont {E.}~\bibnamefont {Lippiello}},
  \bibinfo {author} {\bibfnamefont {A.}~\bibnamefont {Mukherjee}}, \bibinfo
  {author} {\bibfnamefont {S.}~\bibnamefont {Puri}}, \ and\ \bibinfo {author}
  {\bibfnamefont {M.}~\bibnamefont {Zannetti}},\ }\href@noop {} {\bibfield
  {journal} {\bibinfo  {journal} {Phys. Rev. E}\ }\textbf {\bibinfo {volume}
  {85}},\ \bibinfo {pages} {021141} (\bibinfo {year} {2012})}\BibitemShut
  {NoStop}%
\bibitem [{\citenamefont {Paul}\ \emph {et~al.}(2004)\citenamefont {Paul},
  \citenamefont {Puri},\ and\ \citenamefont {Rieger}}]{Paul_2004}%
  \BibitemOpen
  \bibfield  {author} {\bibinfo {author} {\bibfnamefont {R.}~\bibnamefont
  {Paul}}, \bibinfo {author} {\bibfnamefont {S.}~\bibnamefont {Puri}}, \ and\
  \bibinfo {author} {\bibfnamefont {H.}~\bibnamefont {Rieger}},\ }\href
  {\doibase 10.1209/epl/i2004-10276-4} {\bibfield  {journal} {\bibinfo
  {journal} {Europhysics Letters ({EPL})}\ }\textbf {\bibinfo {volume} {68}},\
  \bibinfo {pages} {881} (\bibinfo {year} {2004})}\BibitemShut {NoStop}%
\bibitem [{\citenamefont {Paul}\ \emph {et~al.}(2005)\citenamefont {Paul},
  \citenamefont {Puri},\ and\ \citenamefont {Rieger}}]{ppr05}%
  \BibitemOpen
  \bibfield  {author} {\bibinfo {author} {\bibfnamefont {R.}~\bibnamefont
  {Paul}}, \bibinfo {author} {\bibfnamefont {S.}~\bibnamefont {Puri}}, \ and\
  \bibinfo {author} {\bibfnamefont {H.}~\bibnamefont {Rieger}},\ }\href
  {\doibase 10.1103/PhysRevE.71.061109} {\bibfield  {journal} {\bibinfo
  {journal} {Phys. Rev. E}\ }\textbf {\bibinfo {volume} {71}},\ \bibinfo
  {pages} {061109} (\bibinfo {year} {2005})}\BibitemShut {NoStop}%
\bibitem [{\citenamefont {Berche}(2003)}]{berche2003bulk}%
  \BibitemOpen
  \bibfield  {author} {\bibinfo {author} {\bibfnamefont {B.}~\bibnamefont
  {Berche}},\ }\href@noop {} {\bibfield  {journal} {\bibinfo  {journal} {J.
  Phys. A: Math. Gen.}\ }\textbf {\bibinfo {volume} {36}},\ \bibinfo {pages}
  {585} (\bibinfo {year} {2003})}\BibitemShut {NoStop}%
\bibitem [{\citenamefont {Berthier}\ \emph {et~al.}(2001)\citenamefont
  {Berthier}, \citenamefont {Holdsworth},\ and\ \citenamefont
  {Sellitto}}]{berthier2001nonequilibrium}%
  \BibitemOpen
  \bibfield  {author} {\bibinfo {author} {\bibfnamefont {L.}~\bibnamefont
  {Berthier}}, \bibinfo {author} {\bibfnamefont {P.~C.}\ \bibnamefont
  {Holdsworth}}, \ and\ \bibinfo {author} {\bibfnamefont {M.}~\bibnamefont
  {Sellitto}},\ }\href@noop {} {\bibfield  {journal} {\bibinfo  {journal} {J.
  Phys. A: Math. Gen.}\ }\textbf {\bibinfo {volume} {34}},\ \bibinfo {pages}
  {1805} (\bibinfo {year} {2001})}\BibitemShut {NoStop}%
\bibitem [{\citenamefont {Abriet}\ and\ \citenamefont
  {Karevski}(2004{\natexlab{a}})}]{abriet2004off}%
  \BibitemOpen
  \bibfield  {author} {\bibinfo {author} {\bibfnamefont {S.}~\bibnamefont
  {Abriet}}\ and\ \bibinfo {author} {\bibfnamefont {D.}~\bibnamefont
  {Karevski}},\ }\href@noop {} {\bibfield  {journal} {\bibinfo  {journal} {Eur.
  Phys. J. B}\ }\textbf {\bibinfo {volume} {37}},\ \bibinfo {pages} {47}
  (\bibinfo {year} {2004}{\natexlab{a}})}\BibitemShut {NoStop}%
\bibitem [{\citenamefont {Huse}(1989)}]{huse1989remanent}%
  \BibitemOpen
  \bibfield  {author} {\bibinfo {author} {\bibfnamefont {D.~A.}\ \bibnamefont
  {Huse}},\ }\href@noop {} {\bibfield  {journal} {\bibinfo  {journal} {Phys.
  Rev. B}\ }\textbf {\bibinfo {volume} {40}},\ \bibinfo {pages} {304} (\bibinfo
  {year} {1989})}\BibitemShut {NoStop}%
\bibitem [{\citenamefont {Yeung}\ \emph {et~al.}(1996)\citenamefont {Yeung},
  \citenamefont {Rao},\ and\ \citenamefont {Desai}}]{yeung1996bounds}%
  \BibitemOpen
  \bibfield  {author} {\bibinfo {author} {\bibfnamefont {C.}~\bibnamefont
  {Yeung}}, \bibinfo {author} {\bibfnamefont {M.}~\bibnamefont {Rao}}, \ and\
  \bibinfo {author} {\bibfnamefont {R.~C.}\ \bibnamefont {Desai}},\ }\href@noop
  {} {\bibfield  {journal} {\bibinfo  {journal} {Phys. Rev. E}\ }\textbf
  {\bibinfo {volume} {53}},\ \bibinfo {pages} {3073} (\bibinfo {year}
  {1996})}\BibitemShut {NoStop}%
\bibitem [{\citenamefont {Puri}\ \emph {et~al.}(1991)\citenamefont {Puri},
  \citenamefont {Chowdhury},\ and\ \citenamefont {Parekh}}]{pcp91}%
  \BibitemOpen
  \bibfield  {author} {\bibinfo {author} {\bibfnamefont {S.}~\bibnamefont
  {Puri}}, \bibinfo {author} {\bibfnamefont {D.}~\bibnamefont {Chowdhury}}, \
  and\ \bibinfo {author} {\bibfnamefont {N.}~\bibnamefont {Parekh}},\
  }\href@noop {} {\bibfield  {journal} {\bibinfo  {journal} {Journal of Physics
  A: Mathematical and General}\ }\textbf {\bibinfo {volume} {24}},\ \bibinfo
  {pages} {L1087} (\bibinfo {year} {1991})}\BibitemShut {NoStop}%
\bibitem [{\citenamefont {Puri}\ and\ \citenamefont {Parekh}(1992)}]{pp92}%
  \BibitemOpen
  \bibfield  {author} {\bibinfo {author} {\bibfnamefont {S.}~\bibnamefont
  {Puri}}\ and\ \bibinfo {author} {\bibfnamefont {N.}~\bibnamefont {Parekh}},\
  }\href@noop {} {\bibfield  {journal} {\bibinfo  {journal} {Journal of Physics
  A: Mathematical and General}\ }\textbf {\bibinfo {volume} {25}},\ \bibinfo
  {pages} {4127} (\bibinfo {year} {1992})}\BibitemShut {NoStop}%
\bibitem [{\citenamefont {Mondello}\ and\ \citenamefont
  {Goldenfeld}(1990)}]{mg90}%
  \BibitemOpen
  \bibfield  {author} {\bibinfo {author} {\bibfnamefont {M.}~\bibnamefont
  {Mondello}}\ and\ \bibinfo {author} {\bibfnamefont {N.}~\bibnamefont
  {Goldenfeld}},\ }\href {\doibase 10.1103/PhysRevA.42.5865} {\bibfield
  {journal} {\bibinfo  {journal} {Phys. Rev. A}\ }\textbf {\bibinfo {volume}
  {42}},\ \bibinfo {pages} {5865} (\bibinfo {year} {1990})}\BibitemShut
  {NoStop}%
\bibitem [{\citenamefont {Mondello}\ and\ \citenamefont
  {Goldenfeld}(1992)}]{mg92}%
  \BibitemOpen
  \bibfield  {author} {\bibinfo {author} {\bibfnamefont {M.}~\bibnamefont
  {Mondello}}\ and\ \bibinfo {author} {\bibfnamefont {N.}~\bibnamefont
  {Goldenfeld}},\ }\href {\doibase 10.1103/PhysRevA.45.657} {\bibfield
  {journal} {\bibinfo  {journal} {Phys. Rev. A}\ }\textbf {\bibinfo {volume}
  {45}},\ \bibinfo {pages} {657} (\bibinfo {year} {1992})}\BibitemShut
  {NoStop}%
\bibitem [{\citenamefont {Blundell}\ and\ \citenamefont
  {Bray}(1994)}]{PhysRevE.49.4925}%
  \BibitemOpen
  \bibfield  {author} {\bibinfo {author} {\bibfnamefont {R.~E.}\ \bibnamefont
  {Blundell}}\ and\ \bibinfo {author} {\bibfnamefont {A.~J.}\ \bibnamefont
  {Bray}},\ }\href {\doibase 10.1103/PhysRevE.49.4925} {\bibfield  {journal}
  {\bibinfo  {journal} {Phys. Rev. E}\ }\textbf {\bibinfo {volume} {49}},\
  \bibinfo {pages} {4925} (\bibinfo {year} {1994})}\BibitemShut {NoStop}%
\bibitem [{\citenamefont {Abriet}\ and\ \citenamefont
  {Karevski}(2004{\natexlab{b}})}]{abriet2004off3d}%
  \BibitemOpen
  \bibfield  {author} {\bibinfo {author} {\bibfnamefont {S.}~\bibnamefont
  {Abriet}}\ and\ \bibinfo {author} {\bibfnamefont {D.}~\bibnamefont
  {Karevski}},\ }\href@noop {} {\bibfield  {journal} {\bibinfo  {journal} {Eur.
  Phys. J. B}\ }\textbf {\bibinfo {volume} {41}},\ \bibinfo {pages} {79}
  (\bibinfo {year} {2004}{\natexlab{b}})}\BibitemShut {NoStop}%
\bibitem [{\citenamefont {Paul}\ \emph {et~al.}(2007)\citenamefont {Paul},
  \citenamefont {Schehr},\ and\ \citenamefont {Rieger}}]{paul2007superaging}%
  \BibitemOpen
  \bibfield  {author} {\bibinfo {author} {\bibfnamefont {R.}~\bibnamefont
  {Paul}}, \bibinfo {author} {\bibfnamefont {G.}~\bibnamefont {Schehr}}, \ and\
  \bibinfo {author} {\bibfnamefont {H.}~\bibnamefont {Rieger}},\ }\href@noop {}
  {\bibfield  {journal} {\bibinfo  {journal} {Phys. Rev. E}\ }\textbf {\bibinfo
  {volume} {75}},\ \bibinfo {pages} {030104} (\bibinfo {year}
  {2007})}\BibitemShut {NoStop}%
\bibitem [{\citenamefont {Huse}\ and\ \citenamefont {Henley}(1985)}]{hh85}%
  \BibitemOpen
  \bibfield  {author} {\bibinfo {author} {\bibfnamefont {D.~A.}\ \bibnamefont
  {Huse}}\ and\ \bibinfo {author} {\bibfnamefont {C.~L.}\ \bibnamefont
  {Henley}},\ }\href {\doibase 10.1103/PhysRevLett.54.2708} {\bibfield
  {journal} {\bibinfo  {journal} {Phys. Rev. Lett.}\ }\textbf {\bibinfo
  {volume} {54}},\ \bibinfo {pages} {2708} (\bibinfo {year}
  {1985})}\BibitemShut {NoStop}%
\end{thebibliography}%

\end{document}